\documentclass[rsi,numerical,reprint,amsmath,amssymb,amsfonts,aip,showpacs,nolongbibliography]{revtex4-1}

\usepackage{graphicx}								

\usepackage{dcolumn}								

\usepackage{amsmath}

\usepackage[alsoload=hep]{siunitx}	
\usepackage[T1]{fontenc}							
\usepackage[utf8]{inputenc}							
\usepackage{epsfig}									
\usepackage{graphics}								
\usepackage{subfigure}								
\usepackage{esvect}
\usepackage{amsmath,amssymb,amstext,amsfonts}		
\usepackage[alsoload=hep]{siunitx}					
\usepackage{tikz}
\usepackage{pgfplots}

\newunit{\atomicunit}{\textmd{a{.}u{.}}}			

\usepackage{bm}

\begin{document}

\preprint{PRIOC}

\title{Electron and recoil ion momentum imaging with a magneto-optically trapped target}

\author{R.~Hubele}
\affiliation{Max Planck Institute for Nuclear Physics, Saupfercheckweg 1, 69117 Heidelberg, Germany}

\author{M. Schuricke}
\affiliation{Max Planck Institute for Nuclear Physics, Saupfercheckweg 1, 69117 Heidelberg, Germany}

\author{J. Goullon}
\affiliation{Max Planck Institute for Nuclear Physics, Saupfercheckweg 1, 69117 Heidelberg, Germany}

\author{H. Lindenblatt}
\affiliation{Max Planck Institute for Nuclear Physics, Saupfercheckweg 1, 69117 Heidelberg, Germany}

\author{N. Ferreira}
\altaffiliation{present address: Centro Federal de Educa\c{c}\~ao Tecnol\'ogica Celso Suckow da Fonseca (CEFET-RJ), 20271-110 Rio de Janeiro, Brazil}
\affiliation{Max Planck Institute for Nuclear Physics, Saupfercheckweg 1, 69117 Heidelberg, Germany}

\author{A. Laforge}
\altaffiliation{present address: Physikalisches Institut, Universit\"at Freiburg, 79104 Freiburg, Germany }
\affiliation{Max Planck Institute for Nuclear Physics, Saupfercheckweg 1, 69117 Heidelberg, Germany}

\author{E. Br\"uhl}
\altaffiliation{present address: Physikalisch-Chemisches Institut, Universit\"at Heidelberg,
69120 Heidelberg, Germany}
\affiliation{Max Planck Institute for Nuclear Physics, Saupfercheckweg 1, 69117 Heidelberg, Germany}

\author{V.L.B. de Jesus}
\affiliation{Instituto Federal de Educa\c{c}\~ao, Ci\^encia e Tecnologia do Rio de Janeiro (IFRJ),  Rua Lucio Tavares 1045, 
 26530-060  Nil{\'o}polis/RJ, Brazil}

\author{D. Globig}
\affiliation{Max Planck Institute for Nuclear Physics, Saupfercheckweg 1, 69117 Heidelberg, Germany}

\author{A. Kelkar}
\altaffiliation{present address: Indian Institute of Technology Kanpur, Kanpur 208016, India}
\affiliation{Max Planck Institute for Nuclear Physics, Saupfercheckweg 1, 69117 Heidelberg, Germany}
\affiliation{Extreme Matter Institute EMMI, GSI Helmholtzzentrum f\"ur Schwerionenforschung GmbH, Planckstrasse 1, 64291 Darmstadt, Germany}

\author{D. Misra}
\altaffiliation{present address: Tata Institute of Fundamental Research, Colaba, Mumbai 400 005, India}
\affiliation{Max Planck Institute for Nuclear Physics, Saupfercheckweg 1, 69117 Heidelberg, Germany}

\author{K. Schneider}
\affiliation{Max Planck Institute for Nuclear Physics, Saupfercheckweg 1, 69117 Heidelberg, Germany}
\affiliation{Extreme Matter Institute EMMI, GSI Helmholtzzentrum f\"ur Schwerionenforschung GmbH, Planckstrasse 1, 64291 Darmstadt, Germany}

\author{M. Schulz}
\affiliation{Physics Department and LAMOR, Missouri University of Science \& Technology, Rolla, MO 65409, USA}

\author{M. Sell}
\affiliation{Max Planck Institute for Nuclear Physics, Saupfercheckweg 1, 69117 Heidelberg, Germany}

\author{Z. Song}
\altaffiliation{present address: Institute of Modern Physics, Chinese Academy of Sciences, 730000 Lanzhou, China}
\affiliation{Max Planck Institute for Nuclear Physics, Saupfercheckweg 1, 69117 Heidelberg, Germany}

\author{X. Wang}
\altaffiliation{present address: Shanghai EBIT Laboratory, Institute of Modern Physics, Fudan University, Shanghai 200433, China}
\affiliation{Max Planck Institute for Nuclear Physics, Saupfercheckweg 1, 69117 Heidelberg, Germany}

\author{S. Zhang}
\affiliation{Max Planck Institute for Nuclear Physics, Saupfercheckweg 1, 69117 Heidelberg, Germany}

\author{D. Fischer}
\email{fischer@mpi-hd.mpg.de}

\affiliation{Max Planck Institute for Nuclear Physics, Saupfercheckweg 1, 69117 Heidelberg, Germany}

\date{\today}

\begin{abstract}
A reaction microscope (ReMi) has been combined with a magneto-optical trap (MOT) for the kinematically complete investigation of atomic break-up processes. With the novel MOTReMi apparatus, the momentum vectors of the fragments of laser-cooled and state-prepared lithium atoms are measured in coincidence and over the full solid angle. 
The first successful implementation of a MOTReMi could be realized due to an optimized design of the present setup, a nonstandard operation of the MOT, and by employing a switching cycle with alternating measuring and trapping periods. The very low target temperature in the MOT ($\sim$ \SI{2}{\milli\kelvin}) allow for an excellent momentum resolution. Optical preparation of the target atoms in the excited Li $2^2P_{3/2}$ state was demonstrated providing an atomic polarization of close to \SI{100}{\percent}. While first experimental results were reported earlier, in this work we focus on the technical description of the setup and its performance in commissioning experiments involving target ionization in \SI{266}{\nano\metre} laser pulses and in collisions with projectile ions.

\end{abstract}





\maketitle

\tableofcontents

\section{Introduction}

Our knowledge of the structure of matter and our understanding of the interaction and dynamics in few-particle systems is to a large extent based on the systematic observation of scattering processes starting with the pioneering work of Rutherford \cite{Rutherford1911}. Over the decades the experimental techniques were refined by detecting ejected electrons or emitted photons in addition to the scattered particles.
Scattering studies on atomic and molecular systems experienced a further boost with the development of reaction microscopes (ReMi), also referred to as COLTRIMS (cold target recoil ion momentum spectroscopy) \cite{Ullrich03}. With this technique the charged fragments of atomic or molecular targets can be detected in coincidence essentially with  a $4\pi$ solid angle obtaining their momentum vectors and, thus, providing the full kinematical information of scattering events. In the last 20 years, reaction microscopes have been used extensively to study the dynamics of atomic and molecular fragmentation processes in different experimental situations involving charged particle impact\cite{schulz03, PhysRevLett.82.2496}, synchrotron radiation\cite{Weber2004}, intense femto-\ \cite{PhysRevLett.84.447} and attosecond\cite{FischerA} light pulses, as well as free electron lasers\cite{PhysRevLett.98.203001}.

This experimental technique requires gaseous targets at very low temperatures, as otherwise the thermal momentum spread at room temperature would make the recoil ion momentum measurement essentially insensitive to the scattering dynamics. In conventional reaction microscopes low target temperatures are realized employing supersonic expansion in gas jets. This way, noble gas atoms, molecular gases and clusters or, in few experiments, also atomic hydrogen dissociated by microwave radiation \cite{PhysRevLett.103.053201} are provided for collision experiments. However, the momentum resolution achievable with gas jet targets is still limited by the initial temperature. The lowest temperatures are obtained for helium and are typically below \SI{1}{\kelvin} along the jet expansion direction corresponding to a momentum spread of $\Delta p_\parallel\approx 0.2$\,a.u.

With the development of laser cooling \cite{Metcalf} the experimental tools for the preparation of cold atomic samples have been substantially advanced in the last 25 years. Nowadays, laser-cooling techniques are used in a huge number of experiments, e.g.\ for precision spectroscopy \cite{Hinkley13092013}, atom optics \cite{1367-2630-12-6-065014, KetterleNobel}, and ultra low temperature quantum dynamics (recent examples can be found in Ref.\ \onlinecite{RevModPhys.71.1, 0034-4885-75-4-046401, RevModPhys.84.175} and references therein). Here, magneto-optical traps (MOT) are often used e.g.\ to pre-cool atoms for loading dipole or magnetic traps. Compared to gas jets, MOTs do not only allow for much lower temperatures (typically $<$\SI{1}{\milli\kelvin}), they also extend the number of atomic species available as cold gases substantially. Today more than 20 elements can be trapped in MOTs, among others all alkali and alkali earth metals as well as metastable noble gas atoms \cite{RevModPhys.84.175}.
Moreover, atoms trapped in MOTs can easily be prepared in excited states and even polarized due to electronic transitions accessible by visible lasers light \cite{Spinpolarizedtrap}.

The momentum resolved detection of recoiling ions from MOT targets in so-called MOTRIMS apparatuses was achieved already more than fifteen years ago \cite{PhysRevA.56.R4385} and is now employed conventionally in many experiments \cite{PhysRevLett.87.123201, PhysRevLett.87.123202, PhysRevLett.87.123203, blieck:103102, ganjunPRL, SimoneRSI2012}. However, the first successful coincidence measurement of emitted target electrons with recoil ions in a MOTReMi, i.e.\ the combination of a fully equipped reaction microscope with a MOT target, has been reported only very recently \cite{PhysRevLett.109.113202}. The difficulty in realizing such an experiment is to overcome the apparent intrinsic incompatibility connected to the magnetic fields used in reaction microscopes and MOTs, respectively.

In this paper we give detailed insights in the novel experimental technique of the MOTReMi. The experimental setup consisting of the reaction microscope and the MOT target is discussed and commissioning experiments are reported.

\section{Experimental Setup}

The MOTReMi setup described in this paper was developed for experiments at an ion storage ring, the test storage ring TSR at the Max Planck Institute for nuclear physics in Heidelberg \cite{TSRbooklet, TSR91}, in order to study ion-atom collision dynamics \cite{0953-4075-46-3-031001, Renate}. The combination of a reaction microscope and a MOT as well as its operation in an ion storage ring, requires a novel design of each of the individual components.  Even though the MOTReMi setup is primarily designed to be used in an ion storage ring, it can easily be adapted to work with almost any other projectile species.
In the following, the description of the setup will focus on the distinctive features of the MOT and the ReMi. The general techniques involved in the setup will only briefly be summarized as they are well-documented in literature \cite{Ullrich03, Metcalf}.

\subsection{The reaction microscope}
%

\begin{figure}
\centering
\includegraphics[trim= 3cm 0cm 6cm 0cm, width=0.9\linewidth]{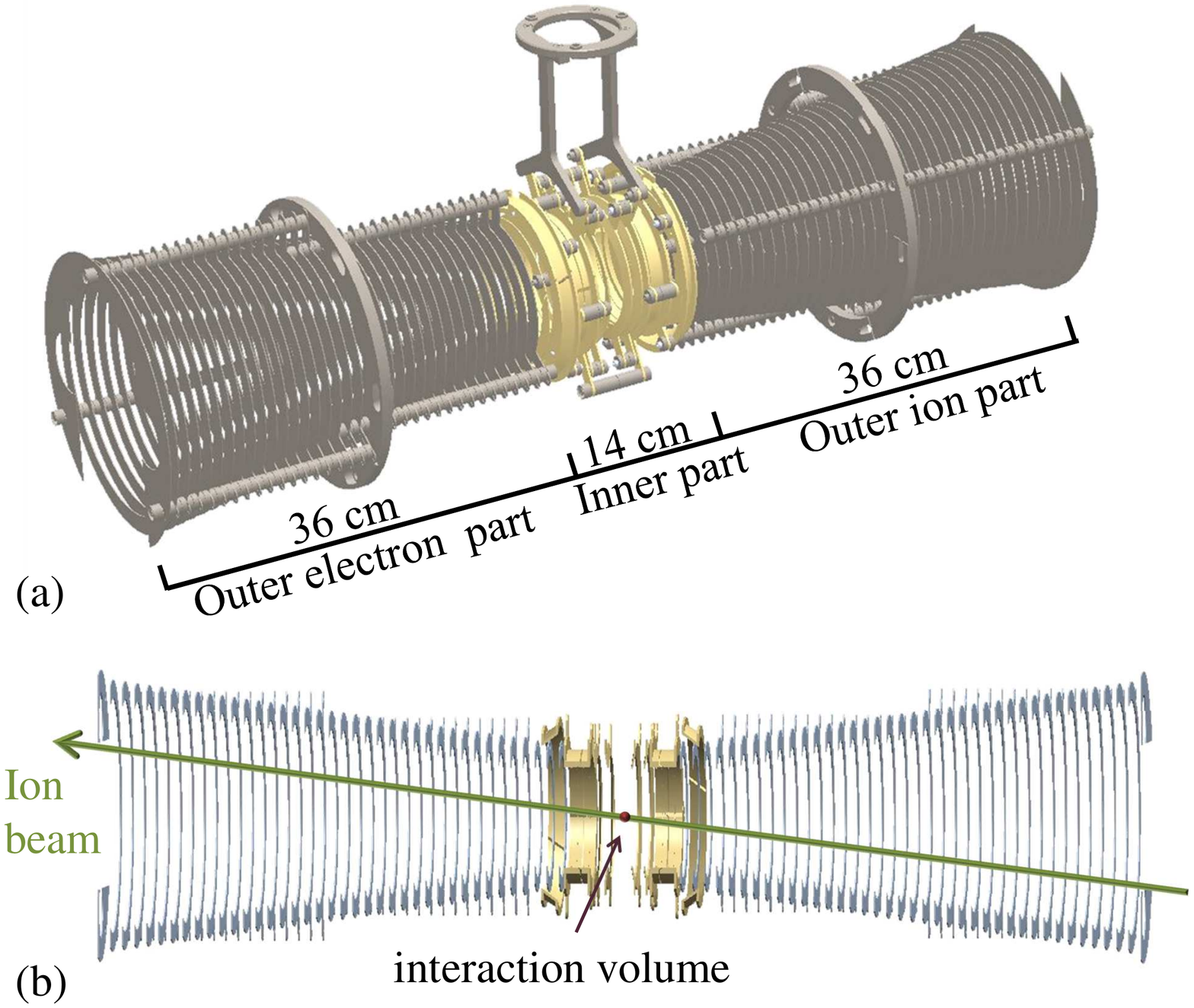}
\caption{Schematic drawing (a) and side view (b) of the spectrometer. It consists of two symmetric outer parts and an inner part with the magnetic MOT coils. The projectile ion beam is indicated by the green arrow in (b).}
\label{spectrometer}
\end{figure}

Reaction microscopes allow for the coincident momentum imaging of the charged fragments of an atomic or molecular target \cite{Moshammer_NuclInstr}. The working principle can be summarized as follows: The target gas and the projectile beam overlap in a well-localized reaction volume within a weak homogeneous electric field. On either side of this extraction field a position- and time-sensitive detector is located, to which electrons and positively charged target ions are guided. The recoiling ions have a kinetic energy of a few ten \SI{}{\milli\electronvolt} or less and can therefore be detected with an effective solid angle of nearly $4\pi$ even with low extraction voltages (typically a few \SI{}{\volt/\centi\meter}). However, emitted electrons usually have considerably higher kinetic energies. In order to increase the electron momentum acceptance, a weak homogeneous magnetic field of a few Gauss is superimposed along the spectrometer axis such that the electrons are guided on spiral trajectories towards the detector. The momenta of the charged fragments can then be calculated from their positions on the detectors and their time-of-flights, which are measured with respect to a reference time given either by a single detected projectile or by a timing signal from a buncher yielding a pulsed beam.

Reaction microscopes are operated in many laboratories worldwide and their mechanical designs as well as their field configurations and imaging properties are adapted to the specific experimental situations. These features are discussed in the following for the present setup.

\subsubsection{Mechanical design of the spectrometer}
\label{mechanicaldesign}

In a reaction microscope, the electrostatic extraction field is generated by electrodes. They have to enable a smooth potential profile avoiding fringe fields close to the particles’ trajectories. Due to the combination of the spectrometer with a magneto-optically trapped target, their design needs to meet additional requirements: 

\begin{itemize}
\item The spectrometer has to accommodate a pair of MOT coils generating a quadrupole magnetic field. As will be detailed in section \ref{RemiinMOT}, the size of these coils should be as small as possible in order to enable a fast switch-off of the field.

\item The target position needs to be optically accessible along three (almost) orthogonal axes for the cooling laser beams which have a diameter of 15 to 20 mm.
\end{itemize}

The implementation of the MOTReMi in the TSR results in an additional constraint:

\begin{itemize} 
\item An open space for the ion beam of $100\times70$ mm in the horizontal and vertical direction, respectively, is required in order to avoid a reduction of the phase space acceptance of the storage ring. Although the aperture can be considerably smaller during data taking, the larger opening facilitates beam adjustment and injection of the initially not yet cooled projectile beam.
\end{itemize}

In general, the spectrometer design depends also on the choice of the extraction direction with respect to the projectile beam axis. In the literature longitudinal as well as transversal extraction schemes are reported \cite{Doerner200095}, both having advantages and drawbacks. The present setup features a close-to-longitudinal extraction with the spectrometer axis being vertically inclined by an angle of 8$\degree$. This geometry was chosen because placing the electron detector downstream of the reaction volume extends the accessible momentum space towards the continuum of the fast moving projectile. There, a non-negligible electron flux occurs in reactions such as e.g.\ ECC \cite{PhysRevLett.99.163201, PhysRevLett.25.1599}(electron capture to the continuum) or projectile ionization \cite{PhysRevLett.88.103202, PhysRevA.84.022707}.

The aforementioned requirements as well as the choice of the extraction direction determine the design of the present spectrometer. A sketch of the setup is shown in Fig.\  \ref{spectrometer}. It consists of 84 ring electrodes and two end caps which are distributed coaxially over a distance of \SI{860}{mm} symmetrically around the reaction volume. Mechanically, the spectrometer consists of three independent parts: The center part accommodates 14 electrodes and the MOT coils, the two outer parts are mirror inverted and comprise 35 rings each.

In order to allow for small MOT coils close to the spectrometer center and for a large projectile beam aperture at the same time, the rings increase in size towards the outside due to the inclination between projectile beam and spectrometer axis. The inner diameters are \SI{100}{\milli\meter} for the central rings and \SI{194}{\milli\meter} at the detector planes. 

Particular attention was paid to the design of the MOT coils (see section \ref{combination}). In all earlier MOTRIMS setups the axes of the coils are oriented perpendicular to the extraction direction. However, the smallest coil size is realized for a coaxial orientation with the spectrometer rings. In the present design, the coils are embedded in the spectrometer and their holders act as electrodes. The coil holders and the innermost ring electrodes are made of gold-plated aluminum which facilitated the fabrication and provides stable electrostatic conditions. All other rings are made of stainless steel. 

Because the MOT magnetic fields are switched during operation, eddy currents could potentially arise from this mode of operation. To avoid such effects, which result in slower magnetic field decays, any closed conducting loops are avoided and all ring electrodes that are exposed to the MOT field are cut.

The position- and time-sensitive detectors consist of stacked \SI{80}{\milli\meter} micro-channel plates (MCP) combined with delay-line anodes for the position readout \cite{Jagutzki2002244}. During data taking, the detectors need to be centrally positioned with respect to the spectrometer axis. In order to increase the storage ring phase space acceptance e.g.\ during ion beam injection, the detectors are mounted on motorized manipulators with a stroke of 50 mm and can vertically be moved with a speed of up to \SI{20}{\milli\metre\per\second} and positioned with a high repeatability.

Three orthogonal pairs of counter-propagating laser beams are used to cool and trap the target atoms. Two of the beam pairs enter the spectrometer from orthogonal directions through a gap of 17 mm between the two central spectrometer rings. In a conventional orthogonal configuration of the laser beams, the direction of the third pair would coincide with the spectrometer axis. In the present setup, however, such a scheme is impossible because this direction is blocked by the particle detectors. Therefore, these beams are horizontally tilted by 12.5$\degree$, passing the detectors on the side.

\subsubsection{Electric and magnetic field configurations}
\label{fieldconfigurations}

\begin{figure}
\subfigure{\includegraphics[trim= 4cm 2cm 3cm 2cm, width=0.45\textwidth]{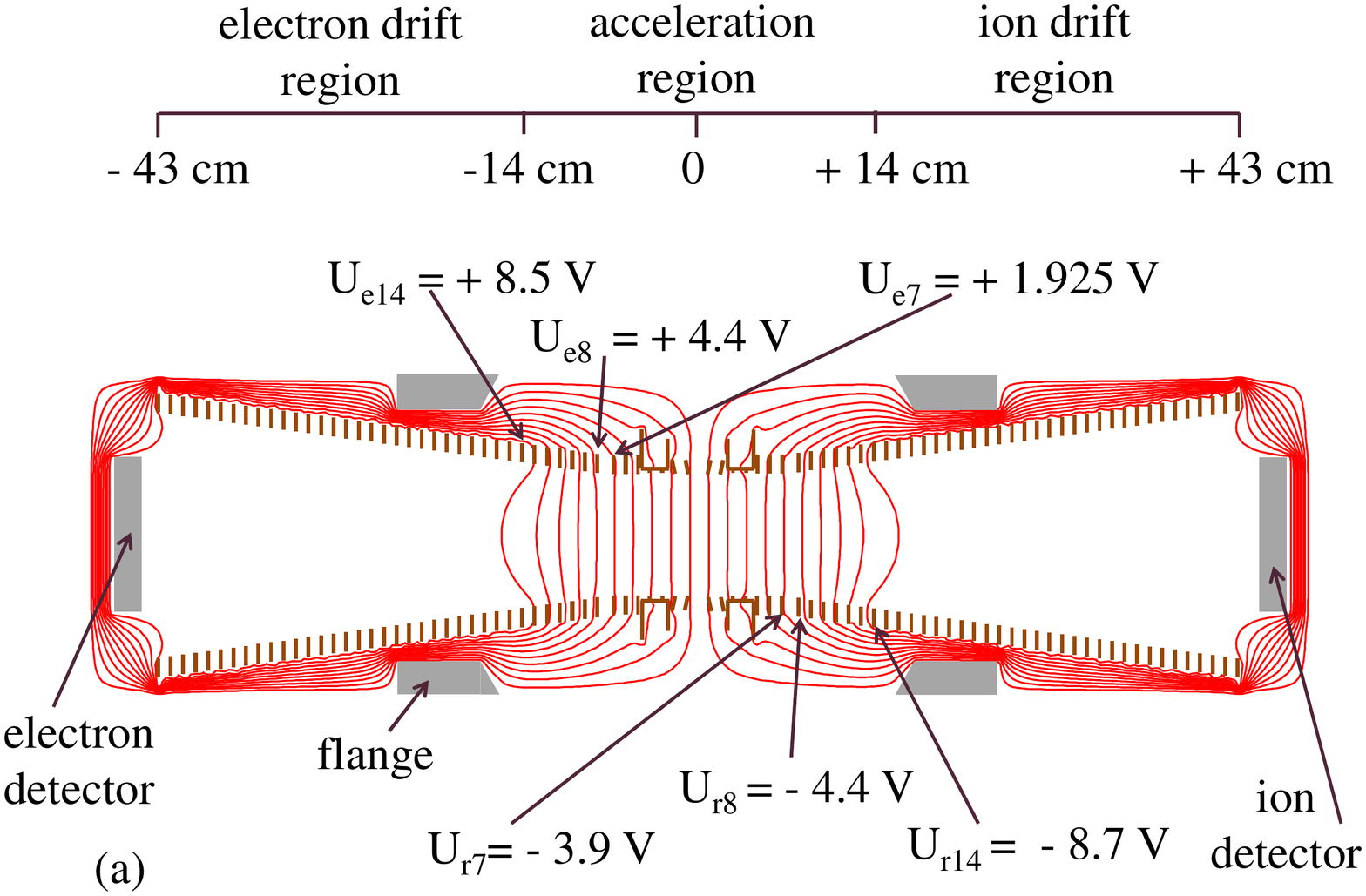}}\hfill \vspace{0.5cm}
\subfigure{\includegraphics[trim= 4cm 4cm 3cm 2cm, width=0.45\textwidth]{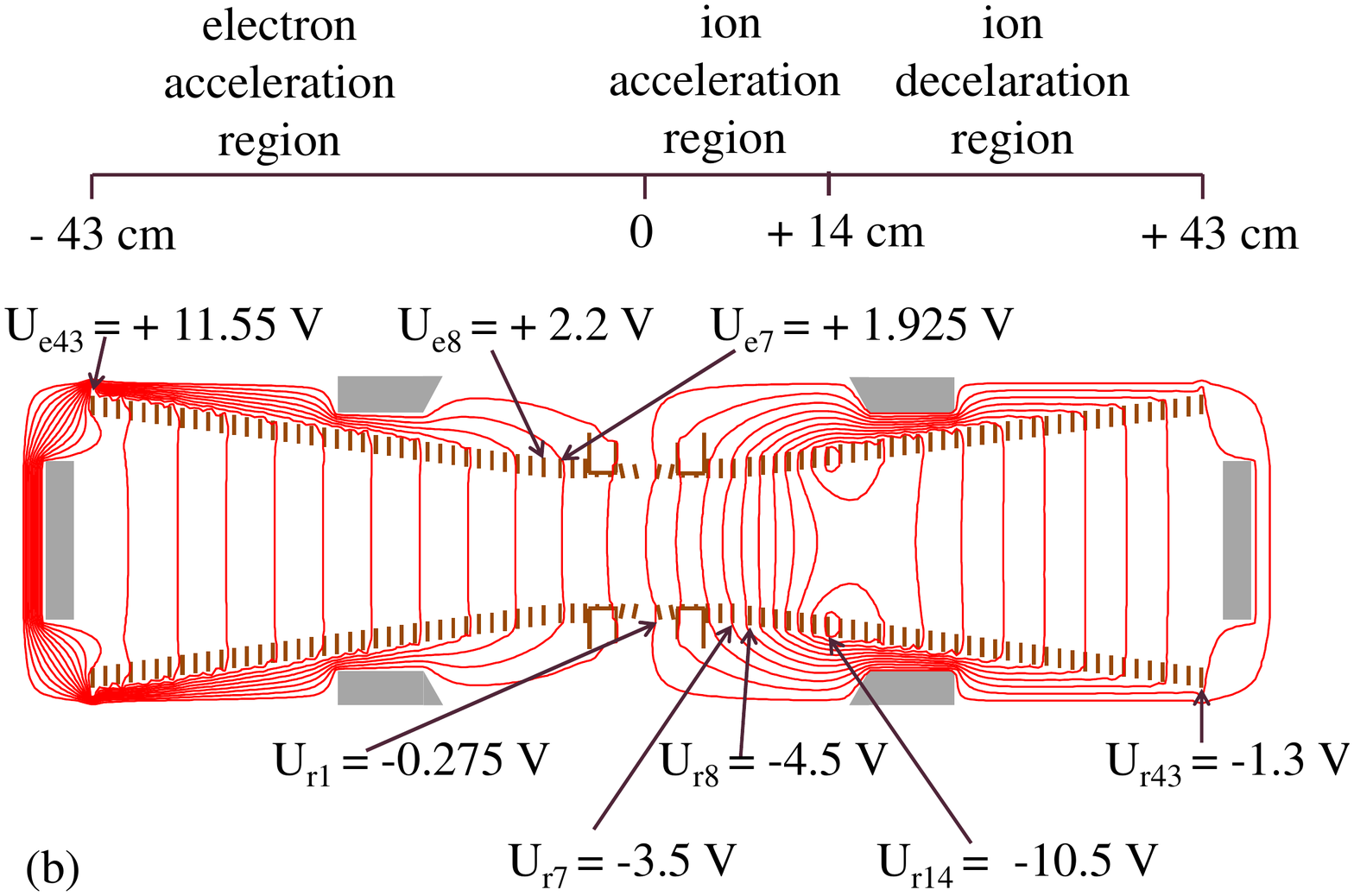}}
\caption{Typical electric field configurations of the spectrometer. Extraction with standard Wiley-McLaren configuration (a) and with spatial focusing for the recoil ions (b) employing an electrostatic lens. Some of the electrodes are connected to external voltage supplies as indicated in the figure. The voltages on all other electrodes are determined by their resistive interconnection (see text).}
\label{fig:Efield_Spectrometer}
\end{figure}

The reconstruction of the momenta relies on smooth and unambiguous imaging properties of the electric extraction field. As mentioned above, the field is generated by 84 ring electrodes each of which needs to be electrically connected to the appropriate voltage source. In order to keep the number of electrical feedthroughs and cables in the vacuum reasonably low, the electrodes are consecutively connected with resistors acting as a voltage divider. The resistors are vacuum prebaked and selected matching the resistance (\SI{100}{\kilo\ohm} and \SI{17}{\kilo\ohm} for the outer and the center part of the spectrometer, respectively) with an accuracy of $\pm 0.1\%$. 16 of the electrodes are directly accessible from the outside of the vacuum chamber allowing for a large flexibility in field configurations.

The ReMi is normally operated in a Wiley-McLaren \cite{Wiley55} or time-focussing configuration. Here the charged fragments are first accelerated in a homogeneous electric field over a distance of $a$ before they pass a field-free drift region of length $d=2a$. With this condition the time of flight of the particles depends to the first order only on their momenta along the extraction direction and not on their initial position. Therefore, the longitudinal momentum resolution is not (or only little) affected by the finite size of the reaction volume. 

With the present experimental setup the condition $d=2a$ is not exactly realizable because a direct adjacency of two regions with constant and zero field, respectively, cannot be generated with ring electrodes. Therefore, the electric fields and the particle trajectories were simulated and optimized using the software SIMION. The field of a standard time focusing configuration is illustrated in Fig.\ \ref{fig:Efield_Spectrometer}(a). It should be noted, that the time-focusing field geometry is of tremendous importance for a high-resolution recoil ion momentum determination, however, for electrons its effect is much weaker and in most cases even negligibly small.

The finite size of the reaction volume also affects the achievable resolution for the momentum components perpendicular to the extraction direction. For the recoil ions, this effect can strongly be reduced by transverse focusing. It is achieved by employing an electrostatic lens close to the reaction volume, where the positions of the ion trajectories are still governed by the ions’ initial coordinates and not by their initial momenta. Even if operating the lens, time focusing can still be realized e.g.\ by increasing the relative length of the drift region compared to the Wiley-McLaren configuration. Such a three-dimensionally focusing spectrometer field has been successfully used and reported earlier (see Ref.\ \onlinecite{Doerner97} and references therein) and it can also be realized with the present setup. 

Alternatively, position and time focusing can simultaneously be achieved by keeping the length of the acceleration region as in the Wiley-McLaren configuration but by changing the drift to a deceleration region. In this configuration, the ions are first accelerated and spatially focused in an increasing electric field before the field reverses, decelerating the ions [see Fig.\ \ref{fig:Efield_Spectrometer}(b)]. The main advantage of this configuration over the standard three-dimensional focusing is that the maximum electric potential in the spectrometer is higher and, as a consequence, the energy acceptance for electrons emitted in direction towards the ion detector is increased. Furthermore, it should be noted, that the benefit of the three-dimensional focusing is limited if electrons and recoil ions are detected in coincidence: First, the operation of the electrostatic lens on the recoil ion side can affect the electron trajectories generally resulting in a reduced resolution of the electron momenta. Second, the magnetic field required for the electron momentum imaging potentially impairs the focusing of the ions. However, a very good recoil ion momentum resolution has still been achieved with this configuration (see section \ref{resolution}).
\\
The magnetic field used to confine the trajectories of the electrons in the transverse direction is generated by a pair of parallel coils with 24 windings each.  The coils are approximately in Helmholtz configuration with $r=$ \SI{85}{\centi\meter} and $D=$ \SI{90}{\centi\meter} ($r$ being the radius of the coils and $D$ their distance). These coils generate a homogeneous magnetic field with variations of less than $0.5\%$ within the spectrometer region. 
In the experiments performed so far field strengths of $B_{0}\leq$ \SI{12}{\gauss} were chosen, corresponding to a current of $I\leq$ \SI{50}{\ampere}. Stronger fields increasingly affect the trapping performance of the MOT.

\subsection{The MOT-Target}
\label{MOT}

The first experimental realization of a trap for neutral atoms exploiting the scattering of near resonant laser radiation was reported in 1987 \cite{PhysRevLett.59.2631}. Since then, magneto-optical traps became a standard tool in experimental physics. A detailed discussion of this technique can be found elsewhere, e.g.\ Ref.\ \onlinecite{Metcalf, PhysRevLett.59.2631}. The working principle is briefly as follows: Atoms exposed to a resonant laser beam experience a force in beam direction by the absorption of photons and their subsequent isotropic re-emission. Illuminating an atom cloud with three pairs of counter-propagating and slightly red-detuned laser beams will result in an effective cooling of the atoms due to the optical Doppler Effect \cite{PhysRevLett.48.596}. Furthermore, in a MOT the resonance frequency of the atomic transition is tuned with an inhomogeneous magnetic field that increases with the distance from the trap center. For an appropriate choice of laser polarizations, this results in a position dependent force towards the trapping center due to the Zeeman shifts of the different magnetic sub-levels of the ground and excited state.
In this chapter, some details of the present setup and its loading system will be discussed. Note, that during the experiments the MOT was operated in a non-standard configuration, which will be detailed in section IIC2.

\subsubsection{The lithium trap}

In the present setup, $^{7}$Li is trapped in the MOT. The choice of this element is motivated by the moderate complexity of lithium as a target system containing only three electrons. This enables to test scattering models on a fundamental level with well-manageable complications due to many electron effects.

Lithium was first trapped in a MOT in 1991 \cite{JJAP.30.L1324}. The optical cooling transition is the D2-line between the $2^{2}S_{1/2}$ ground state and the $2^{2}P_{3/2}$ excited state corresponding to a wavelength of $\lambda$=\SI{671}{\nano\meter}. The ground state splits into two hyperfine sub states with $F=2$ and $F=1$ and a frequency difference of \SI{803}{\mega\hertz}. Therefore, two laser frequencies, the "cooler" and the "repumper", are required in order to maintain a closed cooling transition cycle. In contrast to other alkali metals trapped in MOTs, for lithium the hyperfine-splitting of the excited state is too small (ca.\ \SI{18}{\mega\hertz}) to allow for a selective excitation from the  $^{2}S_{1/2} \,\, (F=2)$ to the $^{2}P_{3/2} \,\, (F=3)$ state. Therefore, a close to symmetric intensity distribution between the cooler and repumper frequencies is required for the efficient trapping of lithium atoms.

In our setup the laser frequencies are provided by a commercial diode laser (Toptica DL pro \cite{Toptica}) which is grating-stabilized to a linewidth of around \SI{1}{\mega\hertz} and subsequently amplified by a tapered amplifier (Toptica TA pro) to a power of about \SI{400}{\milli\watt}. The cooler and repumper beams are prepared by acousto-optical-modulators (AOM) that shift the incoming frequency by several \SI{100}{\mega\hertz} close to the respective resonances. Both beams are mixed and split in a non polarizing beamsplitter and coupled into single-mode fibers. Behind the fibers, they contain cooler and repumper frequencies with intensities typically between \SI{7}{} and \SI{10}{\milli\watt} each.

The laser beams are collimated to a diameter of \SI{15}{} to \SI{20}{\milli\meter} and their polarizations are prepared using combinations of polarizers and $\lambda/4$-plates. The first laser beam is guided and retro-reflected with three mirrors along two perpendicular directions transverse to the axis of the MOT coils, which are used to generate the inhomogeneous magnetic field (details of the coil design are discussed in chapter \ref{RemiinMOT}). Using only one laser beam for the two transversal directions allows operating the MOT with substantially higher laser intensities that would otherwise limit the trapping efficiency. The second beam is used to cool and trap the atoms in the longitudinal direction. It has a small pitch of \SI{12.5}{\degree} with respect to the coil axis (which coincides with the spectrometer axis) thereby passing the electron and recoil ion detectors laterally.

\subsubsection{The 2D MOT loading system}

\begin{figure}
\centering
\includegraphics[angle=0, trim= 1cm 4cm 1cm 2cm, width=0.99\linewidth]{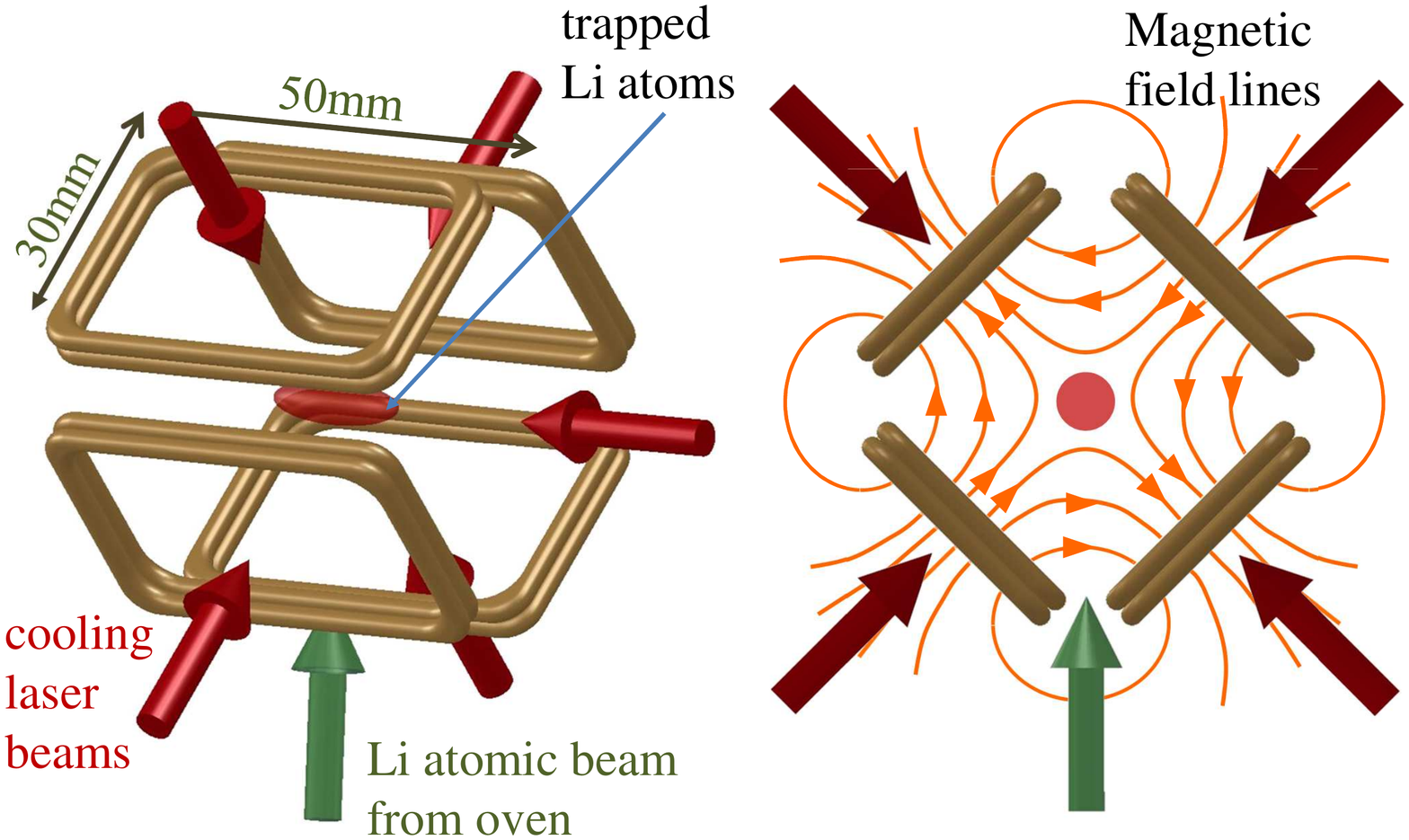}
\caption{Schematic diagram of the 2D MOT.}
\label{fig:2DMOTcoils}
\end{figure}

In order to be trapped in the MOT, lithium atoms need to have a low velocity (a few ten \SI{}{m/s}). In the setup described here, the MOT is loaded with a slow and pre-cooled beam of lithium atoms that was formed in a 2D MOT (e.g.\ as descibed in Ref.\ \onlinecite{PhysRevA.58.3891, Walraven09}). In contrast to Zeeman slowers, which are used in most other setups, in 2D MOTs the atoms are not only slowed down along the atomic beam direction, but also cooled perpendicular to the beam axis. This results in smaller divergence angles of the atomic beam allowing for high capture efficiencies even at relatively long distances between the 2D MOT and the trap. In the present experiment, a large distance between the loading system and the MOT target is required in order to avoid any magnetic stray field in the spectrometer region.

Our 2D MOT setup is based on the design developed by Tiecke et al. \cite{Walraven09}. It is placed at a distance of about \SI{510}{\milli\meter} to the trapping volume. The magnetic quadrupole field is generated by two perpendicular pairs of rectangular Anti-Helmoltz coils, each of which has 4 windings and is formed by a water-cooled copper tube (outer diameter of \SI{3}{\milli\meter}) as depicted in Fig. \ref{fig:2DMOTcoils}. The coils are typically operated at a current of \SI{50}{\ampere} resulting in a magnetic field gradient of ca. \SI{8}{G/\milli\meter}.

For the cooling and trapping in two dimensions only one laser beam is used and guided along two perpendicular axes using three mirrors. The laser beam is collimated to a diameter of about \SI{12}{\milli\meter} and contains cooler and repumper frequencies with intensities of \SI{25}{\milli\watt} to \SI{35}{\milli\watt} each. The lithium oven is situated \SI{160}{\milli\meter} below the 2D MOT trapping region. Here, lithium is heated to temperatures of up to \SI{670}{\kelvin}. Despite the high temperature of the atoms evaporating from the oven, the capture rate in the 2D MOT is still sufficient due to the high laser intensities and magnetic field gradients. 

In the 2D MOT the atoms are cooled and trapped in an ellipsoidal cloud from which they are transferred to the main target MOT with an additional low-intensity (ca. \SI{1}{\milli\watt}) and well-collimated (\SI{3}{\milli\meter} diameter) laser beam. This push beam contains only the repumper frequency which is slightly red-detuned to the $^{2}S_{1/2} \,\, (F=1)$ to $^{2}P_{3/2} \,\, (F=0,1,2)$ transition. After leaving the spatial overlap with the transverse cooling beams, the atoms are pumped into the $F=2$ ground state and do not interact with the push beam any more. In this way a slow directed atomic beam is formed.

\subsection{The combination of  MOT and reaction microscope}
\label{combination}

The biggest challenge in combining the two experimental techniques of reaction microscope and MOT is handling and resolving their intrinsic incompatibility due to different magnetic field configurations: The momentum resolved detection of electrons requires a homogeneous magnetic field, whereas the trapping of atoms in the MOT relies on an inhomogeneous magnetic quadrupole field at the trapping region. 
Generally, this issue is solved by periodically switching between two operation modes, in the first one trapping and cooling the atoms in the MOT, and in the second one acquiring the time and position data of the atomic fragments in the reaction microscope with the quadrupole field being switched off. The switching of the MOT magnetic field was realized already in earlier MOTRIMS experiments in order to allow for a higher recoil ion momentum resolution \cite{blieck:103102, PhysRevLett.87.123202, ganjunPRL}. In the present case, the requirements for the field switching are particularly challenging. On the one hand, the electron trajectories in the spectrometer are very sensitive to field fluctuations which should therefore not exceed about \SI{10}{\milli\gauss}. On the other hand, it is not straight-forward to maintain sufficient target density while the MOT magnetic field is switched off. At a typical temperature of \SI{1}{\milli\kelvin}, the velocity dispersion of the lithium atoms along each coordinate amounts to about \SI{1}{\meter/\second}. For a cloud diameter of \SI{2}{\milli\meter} (FWHM) the number density of the $^{7}$Li atoms drops by a factor of 4 in only one millisecond after switching off the cooling lasers and the magnetic field.
In the following, the specific design and the developed operation scheme for the simultaneous operation of reaction microscope and MOT are described.

\subsubsection{The operation of the reaction microscope: Fast magnetic field switching}
\label{RemiinMOT}

\begin{figure}
\centering
\includegraphics[angle=0, trim= 2cm 1cm 2cm 3cm, width=0.99\linewidth]{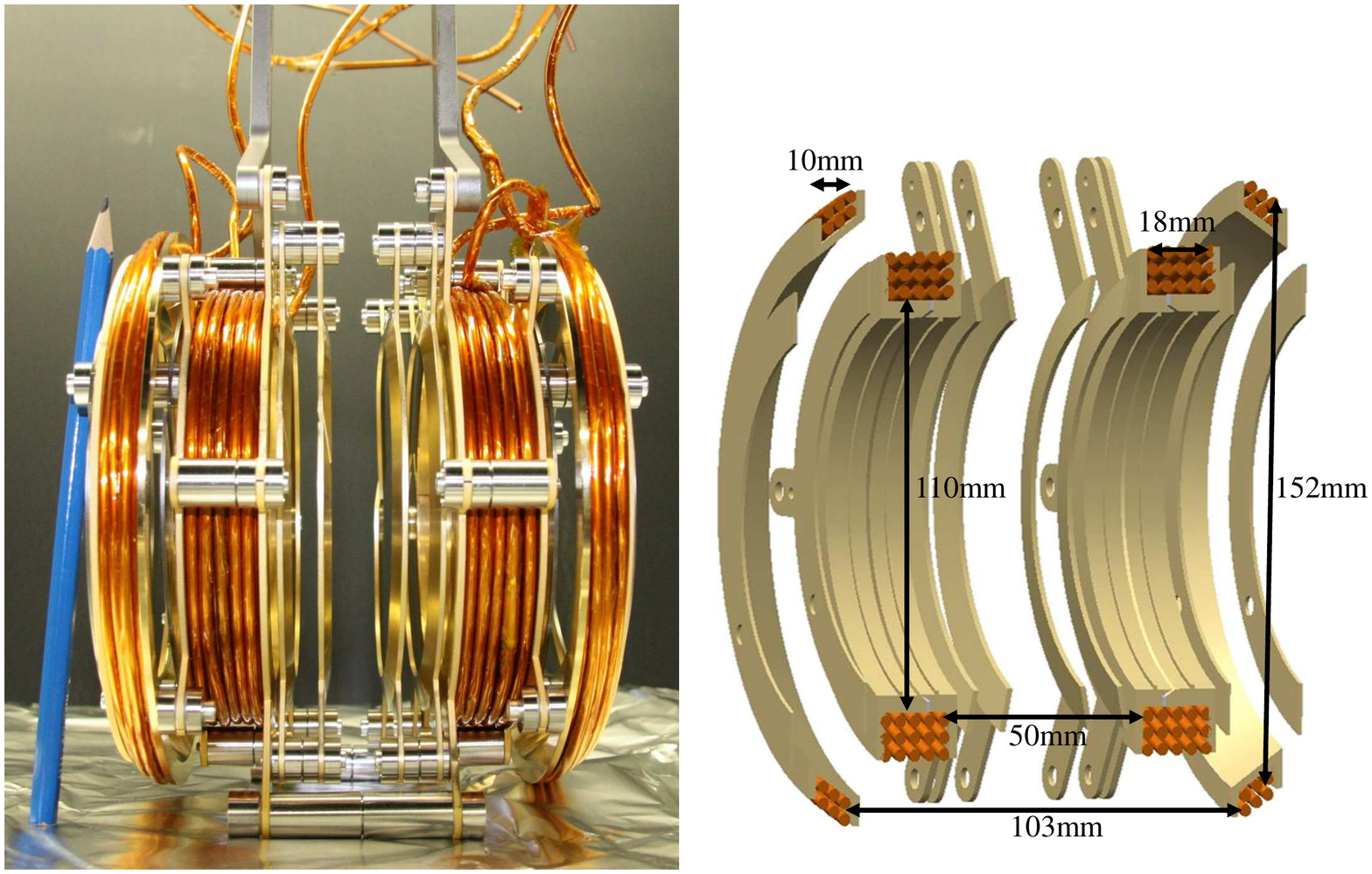}
\caption{Photograph and drawing of the MOT magnetic coils and the inner spectrometer part.}
\label{fig:MOTSpulen}
\end{figure}

Essential for the operation of the reaction microscope, i.e.\ for the momentum resolved detection of the electrons, is the fast switch-off of the MOT magnetic field. In order to achieve a high duty-cycle, a high target density, as well as high electron momentum resolution, the MOT magnetic field should decay to \SI{0.1}{\percent} of its initial magnitude in well below \SI{1}{\milli\second}. Such a switching performance is difficult to realize, because secondary magnetic fields caused by eddy currents induced in the surrounding materials can decay on significantly longer time scales. Therefore, any magnetic flux through conducting loops or metal surfaces (e.g.\ the walls of the vacuum chamber) should be avoided and the MOT magnetic field should be as compact as possible, at the same time providing sufficient magnetic field gradients for the magneto-optical trapping.

The MOT magnetic coils were carefully designed in order to meet the above requirements (see Fig.\ \ref{fig:MOTSpulen}). They are formed by a water-cooled and Kapton-wrapped copper tube (\SI{3}{\milli\meter} outer and \SI{2}{\milli\meter} inner diameter). The size of the coils was minimized by embedding them in the spectrometer ring electrodes. They consist of 15 windings each with an average diameter of \SI{122}{\milli\meter} and a distance of \SI{68}{\milli\meter} from each other. A further confinement of the magnetic field was achieved by introducing compensation coils outside the MOT coils. These coils have a slightly larger diameter (\SI{140}{\milli\meter}) and distance from each other (\SI{120}{\milli\meter}) and they consist of 6 windings each carrying the same current in the opposite direction as their adjacent MOT coils. With this configuration, field gradients of  \SI{9.7}{\gauss/\centi\meter} and \SI{4.5}{\gauss/\centi\meter} at a current of \SI{30}{\ampere} are created in the axial and radial directions, respectively. 

\begin{figure}
 \centering
 \includegraphics[trim= 0cm 0cm 0cm 0cm, width=0.99\linewidth]{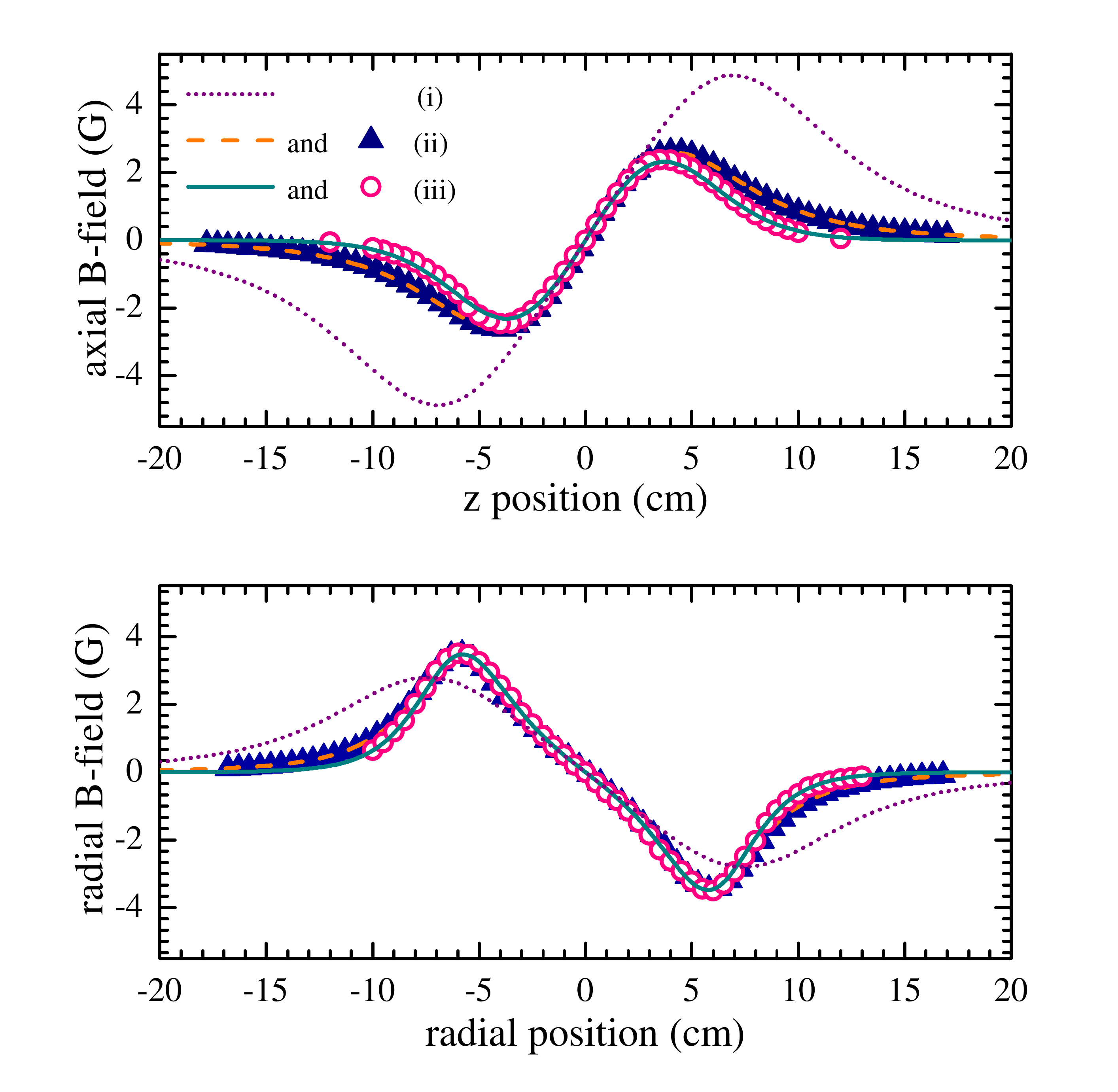}
 \caption{Calculated (lines) and measured (symbols) magnetic field in axial and radial directions for the MOT coil geometries (i), (ii), and (iii) (see text). 
The respective coil currents are  \SI{2.9}{\ampere},  \SI{2.3}{\ampere}, and \SI{3}{\ampere} and adapted to provide similar field gradients at the trap position ($z=r=0$).
}
 \label{fig:Magnetfeldverlauf}
 \end{figure}

\begin{figure}
 \centering
 \includegraphics[trim=4.5cm 2cm 11.5cm 0cm, width=0.8\linewidth]{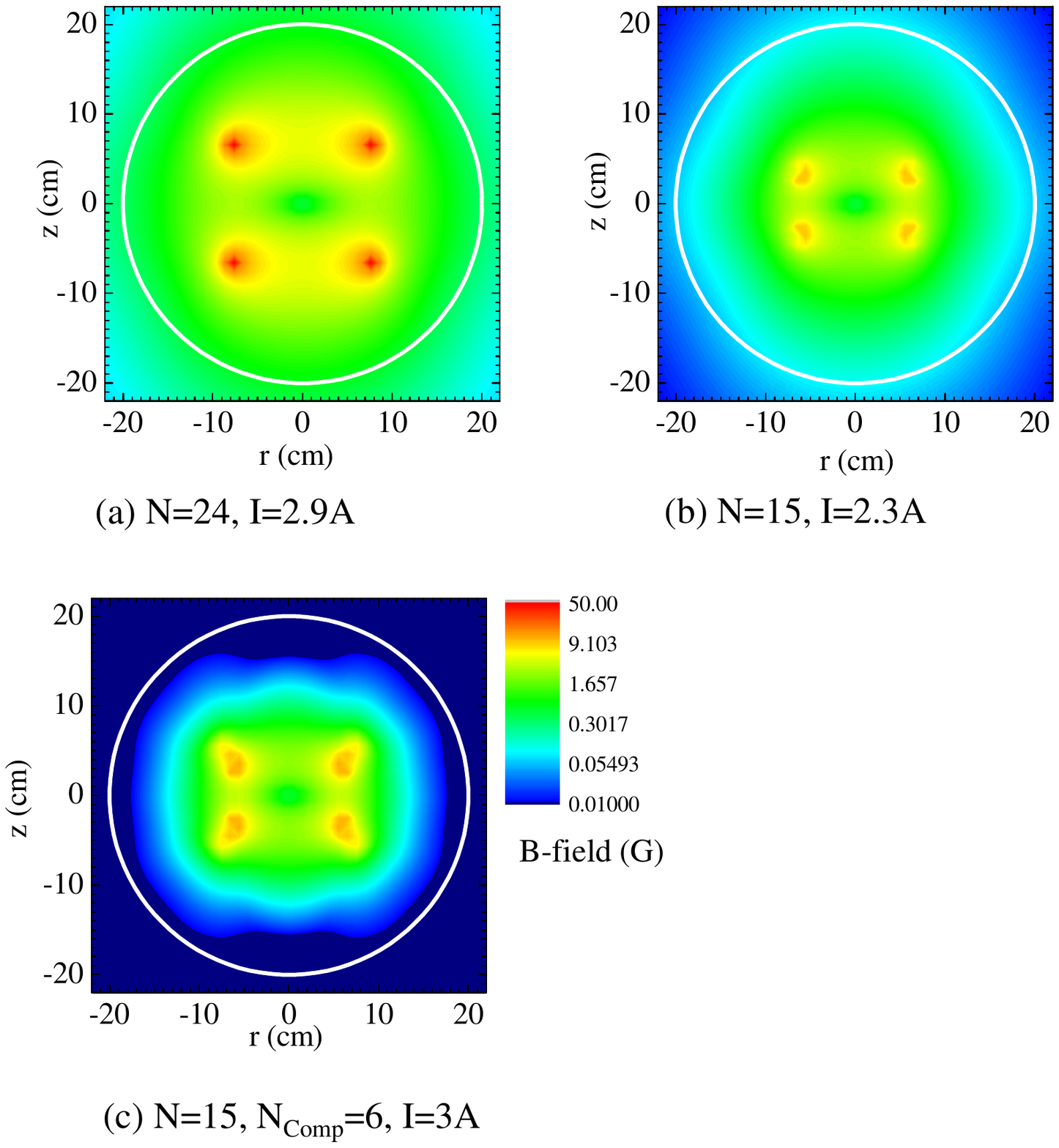}
 \caption{Density plots of the calculated magnetic field strengths for three different designs of the MOT coils. Figure (a) and (b) correspond to the examples (i) and (ii) in the text, respectively. In Fig.\ (c) the field is depicted for configuration (iii) that is used in the present setup. The white circles in the graphs indicate the position of the chamber walls.}
 \label{fig:MOTdensity}
 \end{figure}

The magnetic field was calculated with a MATHEMATICA code from Gehm \cite{Gehmphd} and is in excellent agreement with the measured data along the axial and transversal axes (Fig.\ \ref{fig:Magnetfeldverlauf}). In the figure, three coil configurations are compared: (i) a pair of coils as reported in Schuricke et al. \cite{PhysRevA.83.023413}; (ii) the present MOT coils without compensation coils; and (iii) the standard configuration of the present setup including both MOT and compensation coils. The currents are adapted to provide similar field gradients in the trapping center for all three schemes. The main difference between the configurations (i) and (ii) is the size of the coils. By choosing the minimal size as in (ii), the magnetic field at the walls of the vacuum chamber at a distance of \SI{20}{\centi\meter} from the MOT center is reduced by one order of magnitude compared to (i). Using the compensation coils as in (iii) reduces the field by an additional factor of 10, resulting in a negligible magnetic field strength at the vacuum chamber walls (see also Fig.\ \ref{fig:MOTdensity}).

The switching of the current is performed with a simple metal-oxide semiconductor field-effect transistor MOSFET (IFRB3077). Due to the comparably low inductance of about \SI{70}{\micro H}, the current of \SI{30}{\ampere} can be switched at a time scale of \SI{100}{\micro\second} with moderate corresponding voltages of only about \SI{20}{\volt}.

The decay of the magnetic field can be significantly slower than the switch-off of the coils current. A very sensitive test of the MOT magnetic field decay represents the time-of-flight and position measurement of electrons emitted from the lithium atoms in photoionization processes (see chapter \ref{photoionisation}). After about \SI{250}{\micro\second} only minor changes of the electron trajectories are observed. Small fluctuations occur in the recoil ion time-of-flight about \SI{200}{} to \SI{400}{\micro\second} after the switch-off. They are assigned to mechanical vibrations of the ring electrodes due to the magnetic field switching. However, these small effects are easily accounted for by either choosing appropriate time windows after the switch-off or by using simple correction algorithms in the data analysis.

\subsubsection{Operation of a 2.5D MOT: Trapping in a time-dependent asymmetric magnetic field}
\label{MOTinRemi}


\begin{figure}
\centering
\includegraphics[trim= 4cm 0cm 2cm 0cm, width=0.9\linewidth]{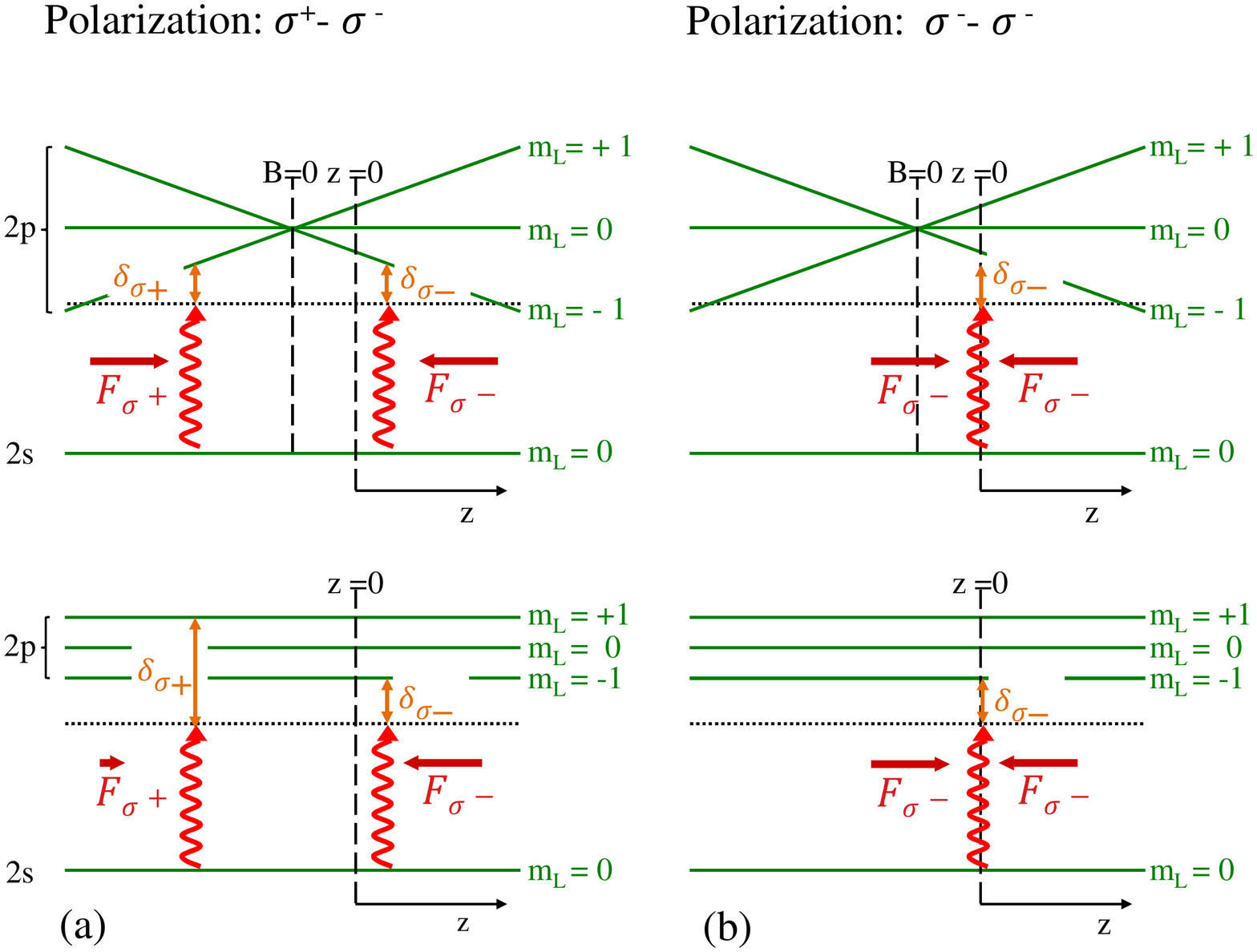}
\caption{Energy level diagram of the 2s and 2p states (neglecting electron and nuclear spins) along the z-Axis. The Zeeman levels are shown for the superposition of MOT and homogeneous ReMi magnetic fields (top row) and for the MOT field being switched off (bottom row). The vertical curvilinear arrows illustrate the absorption of photons in standard MOT (a) and in 2.5D MOT configuration (b) with  $\sigma^{-} -\sigma^{-}$ polarization of the laser light. The horizontal arrows indicate the spontaneous forces due to the interaction of the atoms with the two antiparallel laser beams.}
\label{fig:25DMOTMF}
\end{figure}

\begin{figure}
\centering
\includegraphics[trim= 1cm 3cm 12cm 0cm, width=0.8\linewidth]{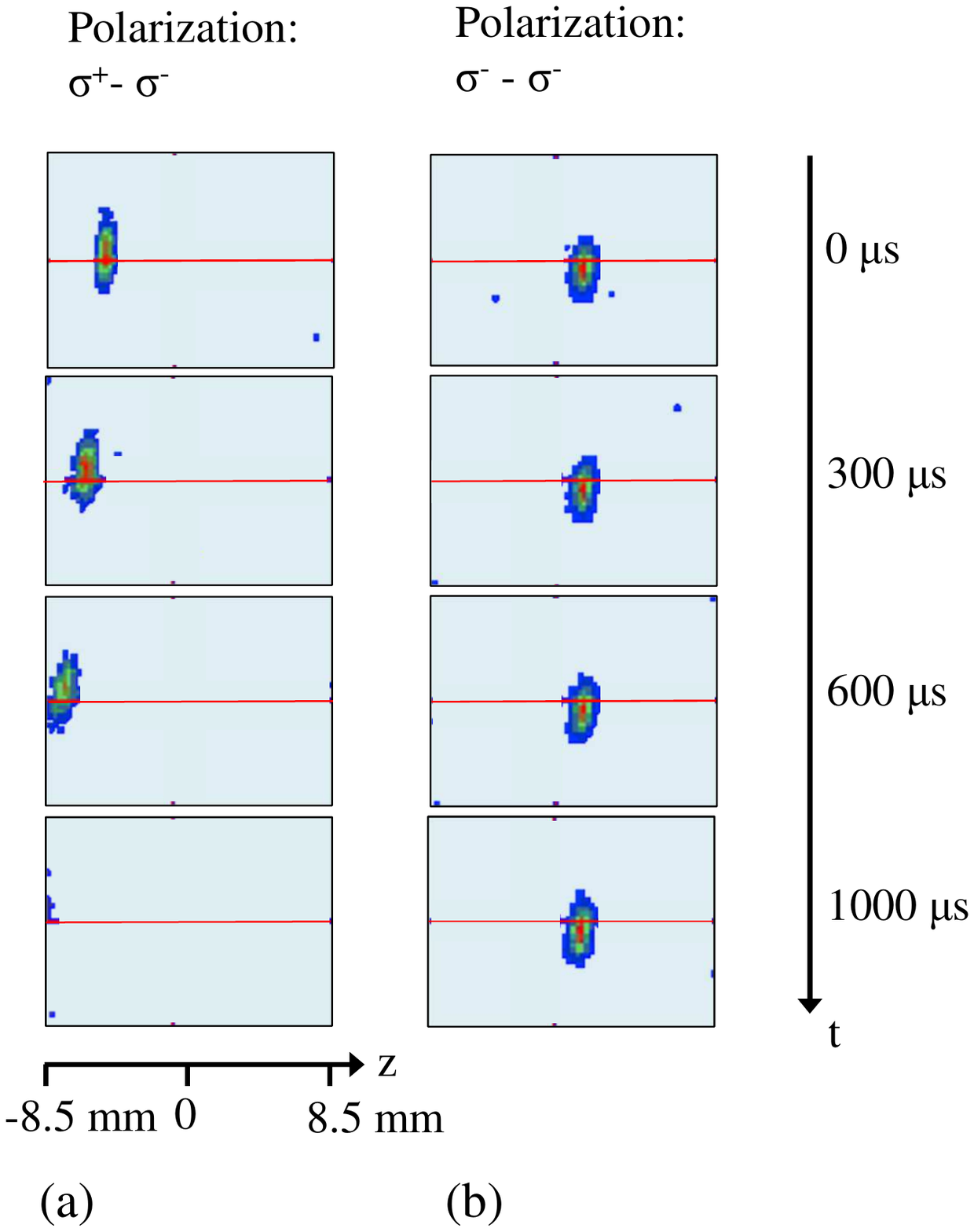}
\caption{Fluorescence images of the lithium cloud taken for several time steps after the MOT magnetic field switch-off with a CCD camera. The homogeneous ReMi magnetic field is oriented along the z direction. In (a) the atoms are trapped in the standard MOT configuration. In (b) the cloud is shown for the 2.5D MOT configuration with $\sigma^{-} -\sigma^{-}$ polarizations of the laser beams along the z direction. The red horizontal line depicts the trap center.}
\label{fig:CCDbild}
\end{figure}

Besides target temperature and density another important property of every MOTRIMS experiment is the obtainable duty cycle, i.e.\ the relative duration of the MOT magnetic field being switched off should be as long as possible. In earlier MOTRIMS experiments with switched magnetic fields \cite{PhysRevLett.103.103008, blieck:103102, PhysRevLett.87.123202} a relatively high duty cycle was obtained because the traps didn't have to be completely reloaded in each switching cycle. Most atoms remained in the vicinity of the trapping position throughout the switching period and could be recaptured with high efficiency. In contrast, the superimposed homogeneous ReMi magnetic field results in the breaking of symmetry of the magnetic field with respect to the center between the MOT coils. This field asymmetry along with the switching of the MOT coils has substantial implications for the trapping and recapturing performance of the MOT.

Conventional magneto-optical trapping of the atoms in the asymmetric magnetic field would result in a displacement of the target cloud from the center between the MOT coils, because for counter-propagating cooling-laser beams of opposite circular polarizations ($\sigma^{+} -\sigma^{-}$) the trapping potential has its minimum at the zero crossing of the magnetic field [Fig.\ \ref{fig:25DMOTMF}(a), top, and Fig.\ \ref{fig:CCDbild}(a), top]. Once the MOT coils are switched off, the homogeneous magnetic field remains resulting in a position-independent Zeeman splitting [Fig.\ \ref{fig:25DMOTMF}(a), bottom], i.e.\ there are different detunings for the $\sigma^{+}$ and $\sigma^{-}$ polarized laser beams. Therefore, the longitudinal forces along the magnetic field direction are not balanced any more pushing the target cloud even farther away from the center [Fig.\ \ref{fig:CCDbild}(a)]. In this conventional mode a measuring period of \SI{1}{\milli\second} was achieved and the trap had to be reloaded in each cycle requiring loading times in the order of seconds. The corresponding duty cycle of less than \SI{0.1}{\percent} could be increased by keeping the atoms in the vicinity of the trapping position e.g.\ by the synchronized switch-off of the laser beams \cite{PhysRevLett.111.133201}. However, without the optical molasses, i.e.\ without the cooling forces of the lasers, the cloud expands more rapidly limiting the maximum duration of the MOT field switch-off.

The above-mentioned secondary effects are avoided by using two laser beams with the same polarizations, (e.g.\ $\sigma^{-} -\sigma^{-}$), along the ReMi magnetic field axis. In this configuration, the longitudinal forces are always balanced making the position of the atom cloud largely independent on the magnetic field [Fig.\ \ref{fig:25DMOTMF}(b)]. However, the laser beams do not exert a trapping force on the atoms in the longitudinal direction which results in a diffusive leaking of the atoms out of the trap. In the present setup, the effective trapping is achieved by a slight misalignment such that the counter-propagating beams are not coaxial but antiparallel with a small offset from each other. The Gaussian intensity profiles of the beams result in a position-dependent scattering rate which, for a proper geometrical beam arrangement, causes the atoms to circulate around the trap center. In this way, there is a coupling of the longitudinal velocities of the atoms to the transverse ones where the MOT cooling and trapping works in a conventional manner and the atoms execute a damped vortex motion.  This trapping configuration, which we dub "2.5D MOT", is reminiscent to earlier reported experiments referred to as "supermolasses" \cite{supermolasses} or as "vortex-force trap"\cite{Spinpolarizedtrap, Walker_vortex_trap}. In this configuration only a minor movement of the atom cloud is observed during the switching cycle [Fig.\ \ref{fig:CCDbild}(b)] and excellent duty-cycles of up to  \SI{50}{\percent} are achieved.


\section{MOT operating parameters and target properties}

The achievable trapping performance of the present 2.5D MOT is, compared to conventional MOTs, much more sensitive to laser beam alignments, polarizations, and detuning. Also other parameters like beam intensities and magnetic field strengths have strong influences on the target density. A systematic and quantitative investigation of the dependences on these parameters is extremely challenging if not impossible, because of the many degrees of freedom, and, even more important, because not all of the parameters can precisely be measured and easily reproduced during the adjustment (e.g.\ beam geometries). In practice, all the parameters are changed and optimized in an iterative process. The achieved target densities and duty-cycles are well reproducible; however, the optimized parameters are not necessarily always exactly identical. In this chapter an overview of typical parameter values and target properties is given.

\subsection{Operating parameters and switching cycle}
\label{switching cycle}

 \begin{table}
  \caption{Typical laser and magnetic field parameters used in the experiment.}
 \centering
 \begin{tabular}{|l|c|c|c|c|}
 \hline Parameter & 2.5D MOT & 2D MOT & push- & excitation\\ 
  &  &  &  beam & beam\\  
  \hline \hline
detunings: &&&&\\
- cooler & \SI{18}{\mega\hertz} & \SI{28}{\mega\hertz} & - & \SI{1}{}/\SI{26}{\mega\hertz} \\ 
- repumper & \SI{15}{\mega\hertz} & \SI{15}{\mega\hertz} & \SI{15}{\mega\hertz} & \SI{15}{\mega\hertz} \\ 
 intensity & \SI{20}{\milli\watt} & \SI{46}{\milli\watt} & \SI{1.5}{\milli\watt} & \SI{5}{\milli\watt}\\ 
  beam diameter & \SI{20}{\milli\meter} & \SI{12}{\milli\meter} & \SI{3}{\milli\meter} &\SI{5}{\milli\meter}\\ 
 \hline 

 \multicolumn{5}{c}{}\\ \hline

\multicolumn{3}{|l|}{ReMi magnetic field strength} & \multicolumn{2}{|l|}{7 - \SI{10}{\gauss}} \\

 \multicolumn{3}{|l|}{2.5D MOT magnetic field gradients:} &\multicolumn{2}{|l|}{} \\

 \multicolumn{3}{|l|}{- longitudinal} & \multicolumn{2}{|l|}{\SI{10}{\gauss/\centi\meter}} \\
 \multicolumn{3}{|l|}{- transversal}& \multicolumn{2}{|l|}{\SI{4.6}{\gauss/\centi\meter}} \\
  \multicolumn{3}{|l|}{2D MOT magnetic field gradient} & \multicolumn{2}{|l|}{\SI{100}{\gauss/\centi\meter}} \\
  \hline 
  \end{tabular}
 \label{MOTPM}
 \end{table}

\begin{figure}
\centering
\includegraphics[trim= 3cm 5cm 3cm 1cm, width=0.9\linewidth]{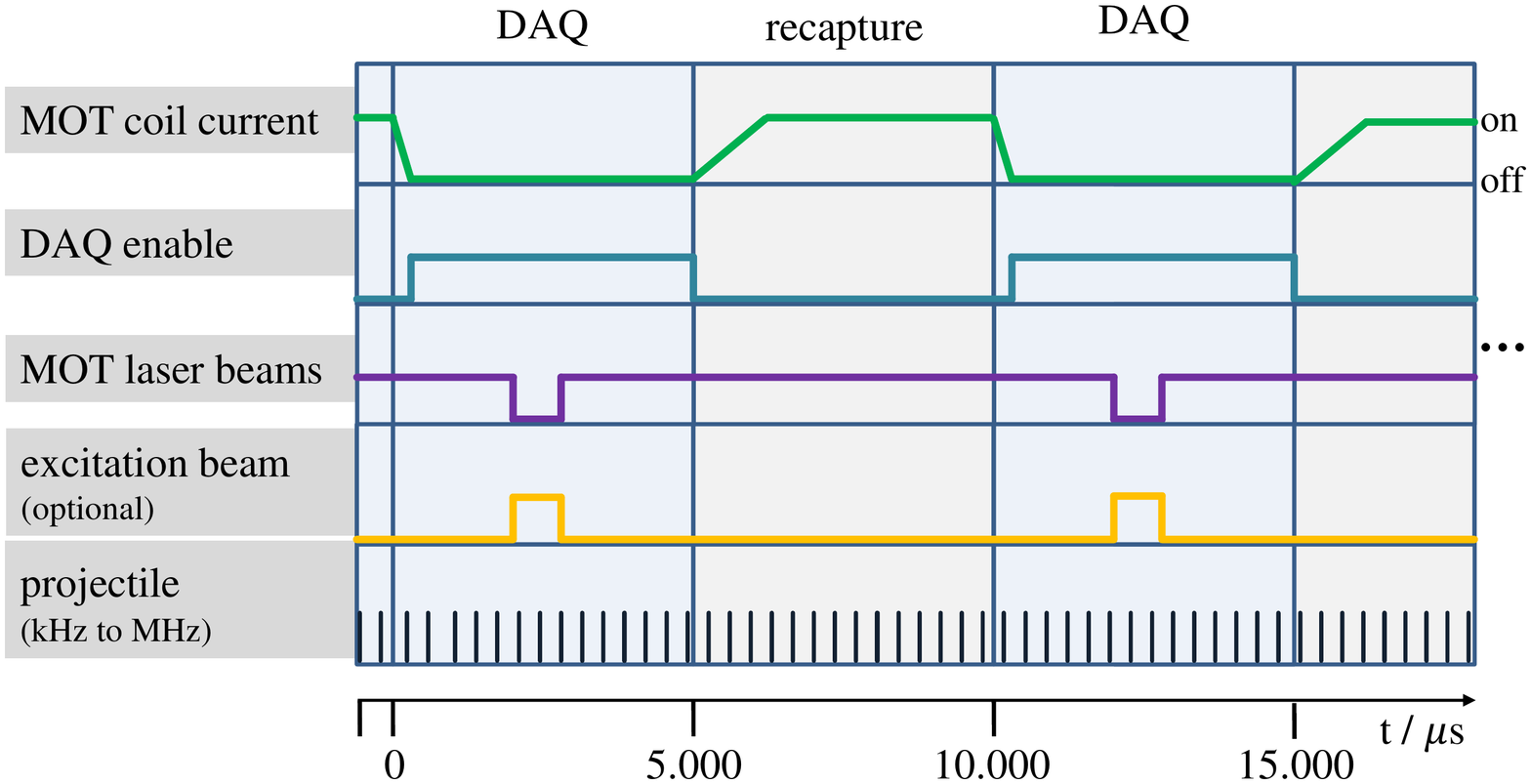}
\caption{Typical switching cycle. Periods of data acquisition (DAQ) and lithium recapture are alternating in a loop. Two full cycles are shown in the graph.} 
\label{dutycycle}
\end{figure}

In table \ref{MOTPM} a set of exemplary MOT operating parameters is given. It should be noted that the parameters are generally correlated. E.g.\ a change of the ReMi magnetic field would make an adaptation of the laser detuning necessary in order to achieve optimal target density. Because such modifications affect the dynamics of the atoms in the trap, a realignment of beam positions (which hardly can be quantified) might be required, too.

After carefully optimizing the parameters, a switching cycle as depicted in Fig.\ \ref{dutycycle} can be executed. The current of the MOT coils is typically switched on and off for durations of \SI{5}{\milli\second}. The data is acquired with the ReMi in a time window that starts about \SI{200}{\micro\second} after the switch-off of the MOT coil current and lasts until the current is switched on again. Within this measuring period the frequency of the cooling lasers can be shifted for about \SI{200}{\micro\second} with an AOM far off resonance. During this time, all the atoms in the target cloud are in the $^2S_{1/2}$ ground state and ionization can be studied state-selectively, whereas during the remaining data acquisition time, a fraction of the atoms (about \SI{20}{\percent}) are in the excited $^2P_{3/2}$ state. Cross sections for the ionization from this excited state can be extracted by calculating the appropriate difference of cross sections measured during time windows when the laser is close to and far off resonance, respectively (cf.\ equation \ref{2pcrosssection}). Optionally, a dedicated cooling laser can be employed during a short time window allowing the optical pumping of the target and selectively populating only one excited magnetic-sublevel (see chapter \ref{polarisation}). In all experiments performed so far, the pulsed projectile beams were not synchronized to the switching cycle of the MOT. However, the repetition frequencies of the projectile pulses are usually much higher (ca.\ \SI{3}{\mega\hertz} for the ion beam in the TSR, \SI{8}{\kilo\hertz} for UV laser system) than the frequency of the MOT switching cycle (\SI{100}{\hertz}) and, therefore, the final event rate is essentially independent on their relative time shifts.

\subsection{Target temperature and density}
\label{Target properties}

The density and temperature of the target gas cloud are decisive parameters for collision experiments because they potentially limit the achievable coincidence rate and momentum resolution, respectively. In the switched mode, the temperature of the trapped atoms does not only influence the resolution but also the expansion of the atomic cloud and therefore the maximum time the MOT magnetic field can be switched off. These parameters were measured by fluorescence imaging where the rate of emitted photons along with the spatial distribution of the target cloud is recorded with a CCD camera.

The scattering rate for the photon absorption and re-emission is given by

\begin{equation}
\Gamma_{SR} = \gamma \rho_{ee} \,.
\label{streurate}
\end{equation}
$\gamma= 1/ \tau$ denotes the inverse lifetime of the excited state and $\rho_{ee}$ is the relative population of the excited state, given by \cite{Metcalf}

\begin{equation}
\rho_{ee} = \dfrac{s_{0}/2}{1+s_{0}+ (2\delta / \gamma)^{2}} \,.
\label{rhoee}
\end{equation}
Here, $\delta$ is the detuning of the laser beams from resonance and $s_{0}= I/I_{s}$ is defined by the ratio of the laser intensity $I$ and the saturation intensity,

\begin{equation}
I_{s} = \dfrac{\pi h c}{3 \lambda^{3} \tau} \,.
\end{equation}
With the laser parameters used in the experiments as shown in table \ref{MOTPM}, the relative population of the excited state is estimated to be $\rho_{ee} =20\%$.

The number of atoms in the trap is obtained by 

\begin{equation}
N_{atom} = \dfrac{N_{ph}}{\Gamma_{SR} \cdot t_{exp}}\,,
\label{Natom}
\end{equation}
where $N_{ph}$ corresponds to the photons emitted during the measuring period $t_{exp}$ (i.e.\ the exposure time of the camera).

Typical atom numbers in the present trap are around \SI{e6}{atoms} in the 2.5D MOT configuration with continuous current at the MOT coils. With atomic cloud diameters of \SI{2}{} - \SI{3}{\milli\meter}, this corresponds to a number density of up to $\rho_{MOT}=$ \SI{e9}{atoms\per\centi\meter^{3}}. In the switched mode this number is reduced by a factor of about 10.

Compared to other lithium MOTs reported in literature \cite{Walraven09, ganjunPRL, GranadeBEC}, the obtained particle numbers are two to three orders of magnitude lower, because the loss rates of atoms in the 2.5D MOT are higher and the loading rates lower as compared to conventional MOT operation. However, in the experiments performed so far the achievable data rate was limited by the performance of the acquisition electronics rather than by the obtainable luminosities because the low target densities could easily be compensated by very high projectile beam intensities.

In general, we do not believe that the presently achieved number densities correspond to a fundamental limit of our operation mode and we expect that there is a large potential of optimizing the trapping performance. This could be achieved e.g.\ by increasing the flux of the atom source or by employing a more elaborate switching cycle where a dedicated period for capturing from the atomic beam could be implemented additionally to the continuous loading of the trap.
\\

The equilibrium temperature of atoms exposed to near-reasonant cooling laser beams is given by \cite{PhillipsNobellecture}

\begin{equation}
k_{B} T = \dfrac{\hbar \gamma}{4} \left(\dfrac{2|\delta|}{\gamma} + \dfrac{\gamma}{2|\delta|} \right) \,
\label{equililbriumtemperature}
\end{equation} 
at sufficiently high laser intensities. For an optimal detuning of $\delta=\gamma/2$ the minimum temperature that can be reached is given by the Doppler cooling limit

\begin{equation}
T_{D} = \dfrac{\hbar \gamma}{2 k_{B}} \,.
\label{equililbriumtemperature}
\end{equation}

In a magneto-optical trap the heating and cooling rates are modified due to the influence of the magnetic fields introducing a velocity dependent scattering force. Therefore, the achieved temperature might be higher as given in equation \ref{equililbriumtemperature} and also - among other factors - depend on the intensities of the laser beams.

For the experimental determination of the MOT temperature a time-of-flight method was used where the expansion of the atomic cloud after the switch-off of magnetic fields and cooling lasers is measured. The atomic cloud expands for a certain period (between \SI{100}{\micro\second} and \SI{900}{\micro\second}) and its size is recorded via fluorescence imaging. The measured size derives from the initial distribution of the cloud and the time-dependent thermal spreading. The temperature estimated in this way is between \SI{2} and \SI{3}{\milli\kelvin}. This is about an order of magnitude higher than the Doppler limit but, corresponding to a momentum spread of the atoms of only \SI{1e-2}{\atomicunit}, is still much smaller than other contributions to the overall momentum resolution (see table \ref{Aufloesung}).

\subsection{State preparation and polarization}
\label{polarisation}

\begin{figure}
\centering
\includegraphics[trim= 2cm 0cm 1cm 0cm, width=0.9\linewidth]{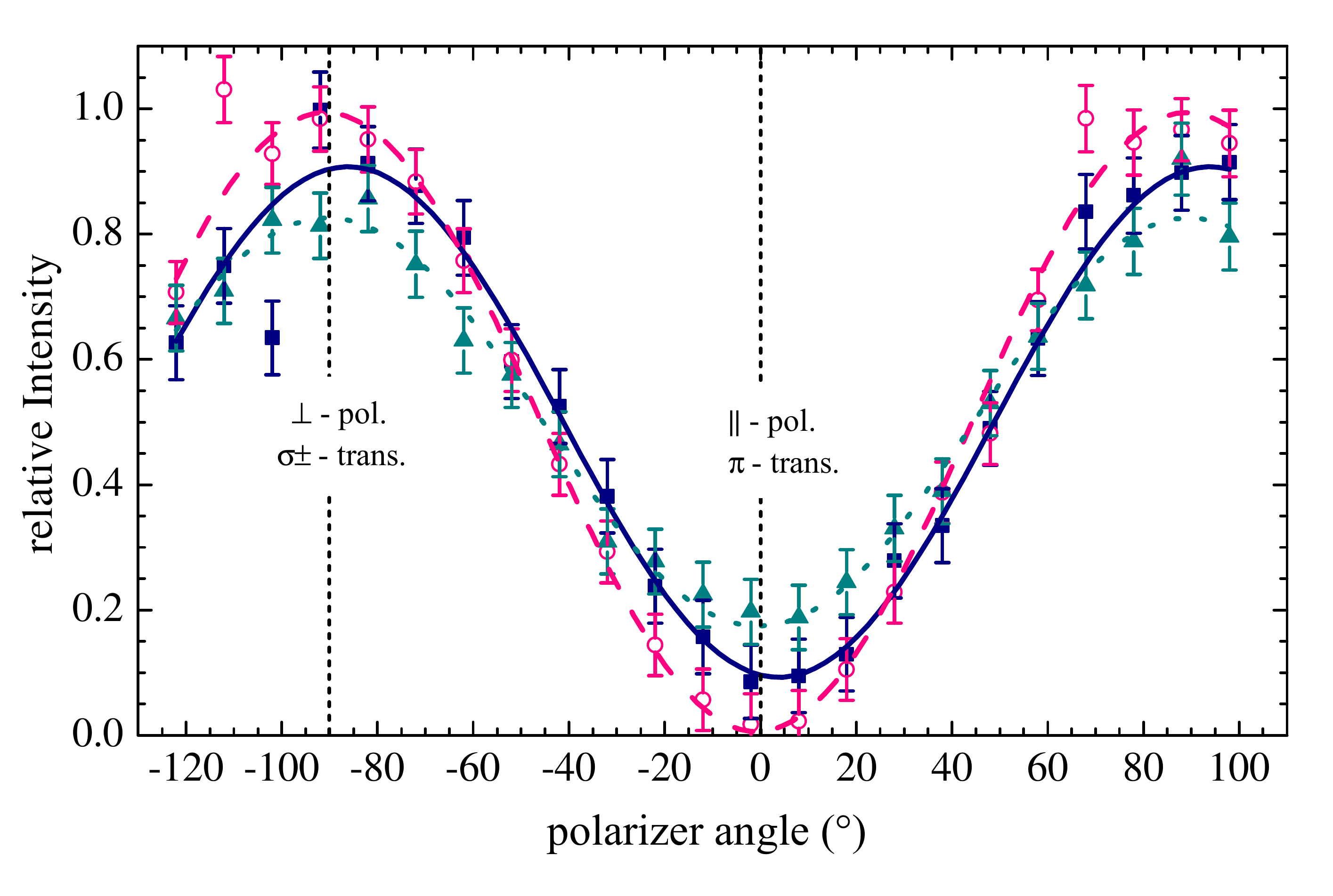}
\caption{Relative intensities of the fluorescence light observed in the direction perpendicular to the magnetic field. A polarizer plate with varying angle is used in front of a CCD camera. The data are for three different excitation schemes: Excitation with the MOT cooling lasers (full triangles), optical pumping with $\sigma^-$ (open circles) and  $\sigma^+$ (full squares) polarized laser beams. The dotted, dashed, and solid lines are the corresponding sinusoidal fits of the measured data.}
\label{Pol_Trans}
\end{figure}

\begin{figure}
\centering
\includegraphics[trim= 2cm 0cm 1cm 0cm, width=0.9\linewidth]{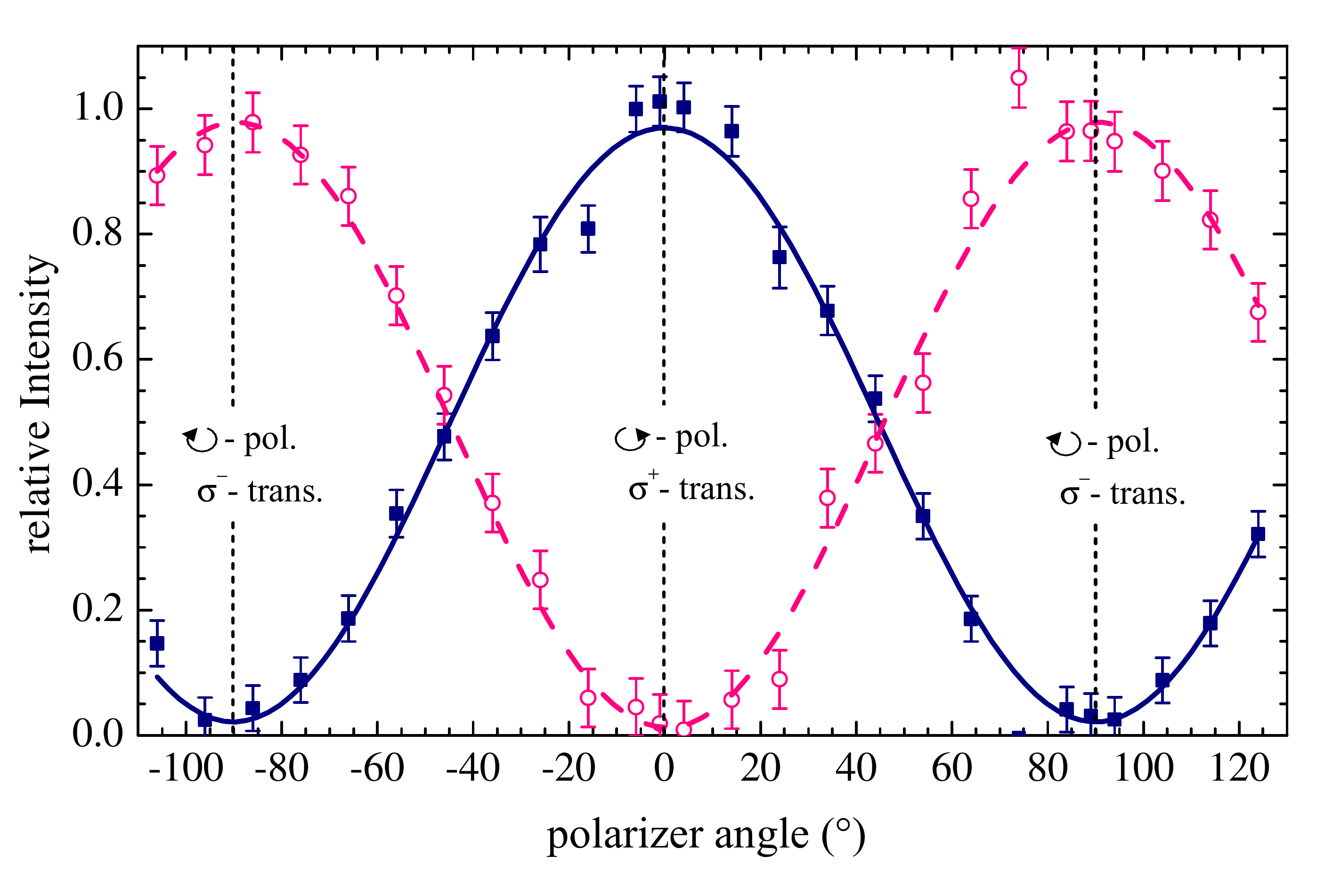}
\caption{As in Fig.\ \ref{Pol_Trans} but the observation angle is 12\degree with respect to the magnetic field direction and a $\lambda/4$ plate is combined with the polarizer in front of the camera. For the excitation scheme including the cooling lasers, the polarization of the fluorescence light has not been analyzed.}
\label{Pol_Long}
\end{figure}

Lithium atoms in the magneto-optical trap are easily prepared in an excited state for scattering experiments, because already the cooling lasers promote a fraction of the atoms ($\rho_{ee}$, cf.\ Eq.\ \ref{rhoee}) to the $2^{2}P_{3/2}$ state. However, the separation of state-selective cross sections for the ionization from the excited state requires the subtraction of the background due to ground state ionization. To this end, the $2^{2}S_{1/2}$ ionization cross sections $\sigma_{2s}$ are measured during a short time period of typically \SI{200}{\micro\second} in the switching cycle when the laser beams are switched off (see chapter \ref{switching cycle}).  The cross sections $\sigma_{2p}$ for the ionization from the excited $2^{2}P_{3/2}$ state are then deduced from the distributions with the lasers being switched on, $\sigma_{2s+2p}$, according to the relation \cite{0953-4075-46-3-031001} 
\begin{equation}
\sigma_{2p} = \dfrac{1}{\rho_{ee}} [\sigma_{2s+2p}-(1-\rho_{ee} )\sigma_{2s}]\,.
\label{2pcrosssection}
\end{equation}

For many experiments it is desirable not only to excite the target atoms but also to prepare them in a specific magnetic sublevel.
Already the red-detuning of the laser beams in combination with the homogeneous magnetic field remaining when the MOT field is switched off result in the polarization of the target atoms, because the lowest lying Zeeman levels of the excited state are predominantly populated (cf.\ Fig.\ \ref{fig:25DMOTMF}). For these levels the electron orbital momentum is mostly oriented anti-parallel to the magnetic field direction (i.e.\ $m_{L}=-1$). 

The degree of polarization can be increased by employing optical pumping \cite{Metcalf}. For that purpose, the cooling laser beams are switched off for about \SI{200}{\micro\second} and the atoms are illuminated by an additional pair of counterpropagating beams almost parallel (with an angle of about 12$\degree$) to the magnetic field direction. The beams contain the cooler frequency, which can be shifted by a few ten MHz with an AOM, and the repumper frequency. By choosing the appropriate detuning and helicity of the light ($\sigma^+$ or $\sigma^-$) the atoms are excited to the $2^{2}P_{3/2}$ state, with either $m_{L}=+1$ or $m_{L}=-1$, respectively. In this way even spin-polarization of the target can be realized \cite{Spinpolarizedtrap}.

The population of the magnetic sublevels was measured by fluorescence imaging, exploiting the different angular emission characteristics and polarizations of the light emitted in $\sigma^+$, $\pi$, and $\sigma^-$ transitions. The intensities of these transitions correspond to the populations of the excited $m_{L}= +1$, $m_{L}= 0$, and $m_{L}= -1$ states, respectively. 
The relative population $P_0$ of the $m_L=0$ state is extracted from the polarization of fluorescence light that is emitted in the direction perpendicular to the magnetic field (see Fig.\ \ref{Pol_Trans}). In this direction, the fraction of the light polarized parallel to the quantization axis (i.e.\ for polarizer angles of $0\degree$ in the figure) originates from $\pi$ transitions. For the observation direction along the quantization axis, there is no intensity from $\pi$ transitions and the ratio between the intensities of left and right circularly polarized light directly reflects the ratio of the populations $P_{+1}$ and $P_{-1}$ in the $m_{L}=+1$ and $m_{L}=-1$ state, respectively. Here, a $\lambda/4$ plate and a polarizer were combined and positioned in front of the camera. In Fig.\ \ref{Pol_Long} the intensity is shown for varying relative angle between $\lambda/4$ and polarizer plate and for an observation angle of 12$\degree$ with respect to the quantization axis.


The populations $P_{-1}$, $P_{0}$, and $P_{+1}$ obtained for the excitation with all the cooling beams and with optical pumping using $\sigma^+$ or $\sigma^-$ polarization are extracted from Figs.\ \ref{Pol_Trans} and \ref{Pol_Long} and listed in table \ref{Polarisationsgrad}. The degree of polarization $\mathcal{P}$ is defined as

\begin{equation}
\mathcal{P}=\sqrt{\dfrac{(P_{-1}-P_{0})^2+(P_{-1}-P_{+1})^2+(P_{0}-P_{+1})^2}{2}}\,.
\end{equation}

For the optical pumping with the $\sigma^-$ polarized excitation beam a polarization degree close to 100\,\% is achieved. For $\sigma^+$ excitation, the optical pumping is less efficient because in this case the laser frequency is near-resonant for transitions to several Zeeman levels. Therefore, many Zeeman levels are excited and the optical pumping efficiency is diminished. However, even for the $\sigma^+$ excitation a polarization of about \SI{90}{\percent} is achieved.

\begin{table}
\centering
\begin{tabular}{|l|c|c|c|c|}
 \hline & $\ P_{-1}\ $& $\ \ P_{0}$\ \ &\ $P_{+1}$\ &$\ \ \mathcal{P}$\ \  \\ 
\hline 
\hline $\sigma^-$ excitation & $0.99$ &  $0.00$ & $0.01$ &  $0.98$\\ 
$\sigma^+$ excitation & $0.02$ & $0.05$ & $0.93$ &$0.90$\\ 
exc.\ with cooling lasers &0.86\footnotemark[1] & 0.09& 0.05\footnotemark[1] &\\ 
\hline \hline
\end{tabular} 
\caption{Relative populations $P_{-1,0,+1}$ of the magnetic sublevels and degree of polarization $\mathcal{P}$ for different excitation schemes. The experimental errors are about 0.02. }
\label{Polarisationsgrad}
\footnotetext[1]{estimated values}

\end{table}

\section{Commissioning experiments}
\label{commissioning}

The novel MOTReMi apparatus was used in several experimental runs studying ion-atom collision dynamics. Experimental cross sections as well as comparisons to theory and interpretations are published elsewhere \cite{PhysRevLett.109.113202, 0953-4075-46-3-031001, Renate}. Here, experimental data of photoionization and ion-impact ionization will be discussed with the focus on the momentum imaging characterization of the setup and the analyses of the data.

\subsection{Photoionization}
\label{photoionisation}

The ionization of target atoms by single photons with well-defined energy and polarization is an ideal process to characterize the properties of the momentum spectrometer. For photoionization the energy and momentum conservation results in particularly simple conditions for the electron and recoil ion momenta $\vec{p_{e}}$ and $\vec{p_{r}}$:

\begin{eqnarray}
\vec{p_{e}}&=&-\vec{p_{r}}\label{photoionisation_relation1}\\
|p_{r}| = |p_{e}| &=& \sqrt{2m_{e}(\hbar \omega-I_{P})}
\label{photoionisation_relation2}
\end{eqnarray}

There, the target ionization potential is $I_{P}$, the photon energy is $\hbar \omega$, and the photon momentum is neglected. Moreover, for some geometries the angular emission characteristics can easily be deduced due to dipole selection rules.

In the present experiment a compact Nd:YAG laser system, a so-called MicroChip laser (Teem Photonics Inc.), is integrated in order to perform calibration measurements before and after each experimental run. The laser is passively Q-switched providing pulse lengths of \SI{600}{\pico\second} with repetition rates of about \SI{7}{\kilo\hertz}. The wavelength of \SI{266}{\nano\meter} is obtained by quadruplication of the Nd:YAG fundamental frequency and corresponds to a photon energy of \SI{4.66}{\electronvolt}.

Because the ionization threshold of the $2^2S_{1/2}$ state is \SI{5.39}{\electronvolt}, the atoms cannot directly be ionized by absorbing a single photon. However, atoms prepared in the excited $2^2P_{3/2}$ state with $I_P=$\SI{3.54}{\electronvolt} are ionized with a high probability resulting in a kinetic energy of the ejected electrons of $\hbar \omega - I_{P}=$\SI{1.12}{\electronvolt}.

\subsubsection{Momentum reconstruction and resolution}
\label{resolution}

\begin{figure}
\subfigure{\includegraphics[width=0.24\textwidth]{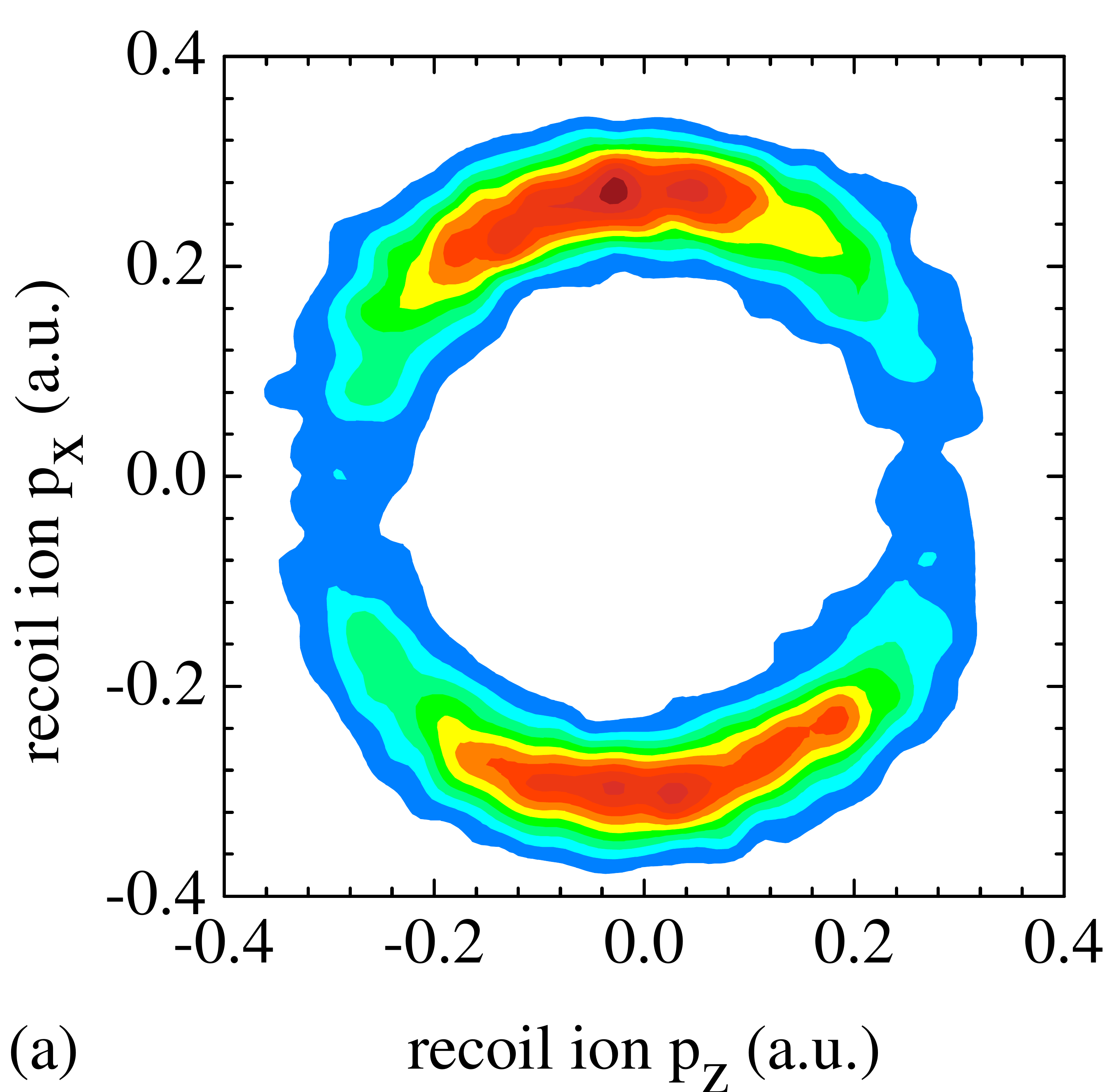}}\hfill
\subfigure{\includegraphics[width=0.24\textwidth]{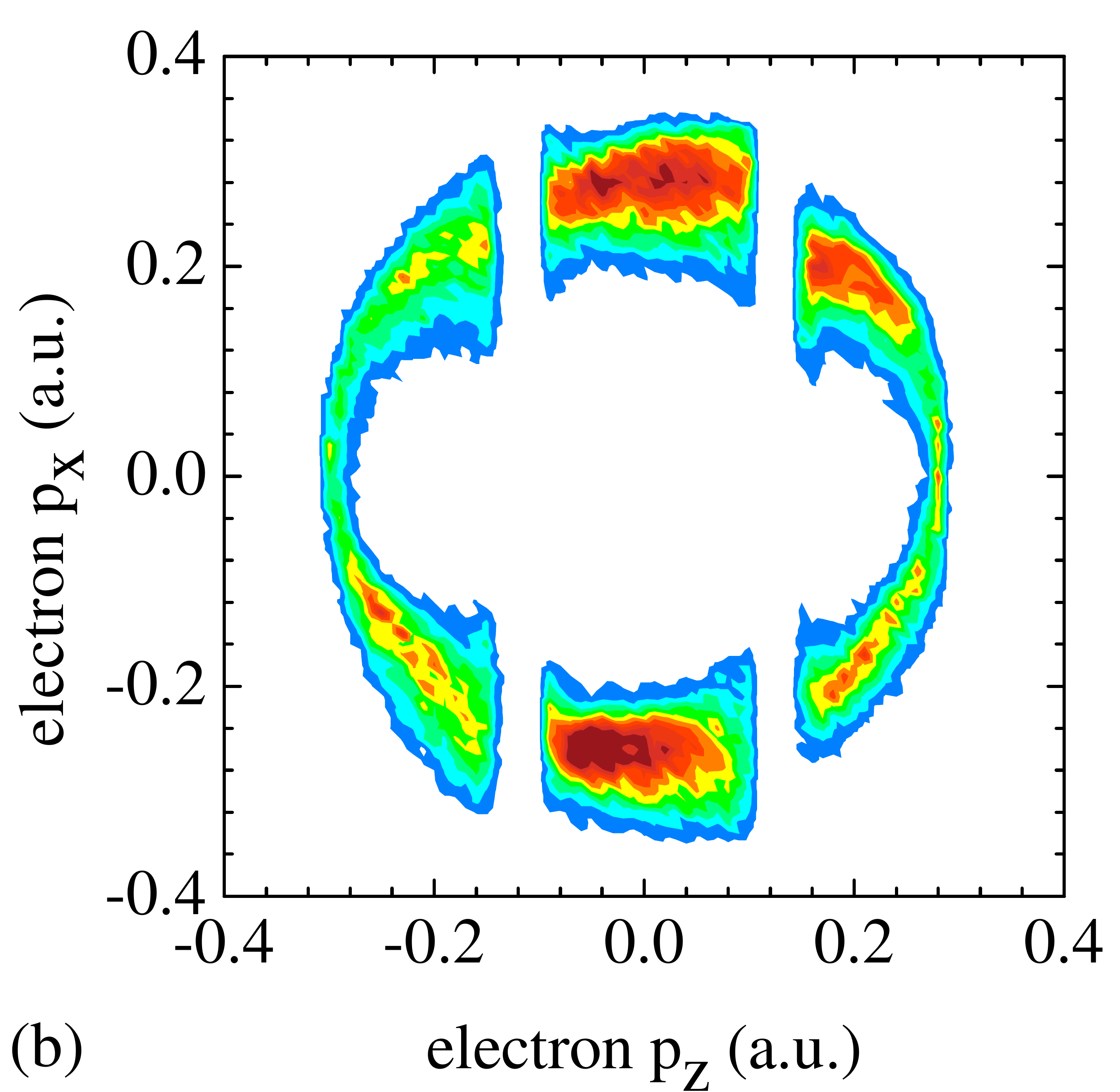}}
\caption{Typical Li$^+$ ion (a) and electron (b) momentum distributions in the xz-plane after photoionization from the $2^2P_{3/2}$ state in a \SI{266}{\nano\metre} laser pulse.}
\label{pxz}
\end{figure}

Calibrating the measured spectra can be very challenging particularly for ion-impact ionization, because the momentum distributions are often broad and do not feature significant and narrow structures or resonances. Therefore, the accurate knowledge of the spectrometer fields and geometry is indispensable for the calibration \cite{Buch_Mos}. For photoionization, however, this task is significantly simplified due to the relation of momenta and energy given in Eqs.\ \ref{photoionisation_relation1} and \ref{photoionisation_relation2}. Here, electrons and target ions are emitted back-to-back and their momenta have to have a constant absolute value, i.e.\ they are on a sphere in momentum space with a well-defined radius.

In the present experiment, the excess energy is \SI{1.12}{\electronvolt}, corresponding to a momentum of \SI{0.286}{\atomicunit} for both fragments. In Fig.\ \ref{pxz} typical recoil ion (a) and electron (b) momentum distributions are shown in the xz-plane. Here, the z direction corresponds to the spectrometer axis while the x direction can be any axis perpendicular to it. Both distributions feature a ring-like structure with a radius of \SI{0.286}{\atomicunit} Additionally, the electron momentum spectrum exhibits two vertical stripes with no intensity. There, all the electrons perform an integer number of full cyclotron revolutions hitting the detector on the same spot (see chapter \ref{wiggles} for details). For these electrons, no information on their transverse momentum can be obtained, and the data is therefore cut out in the analysis.

The resolution of the momentum components can easily be extracted from the width of the ring structures in the spectra in Fig.\ \ref{pxz}. The best achieved resolutions in photoionization measurements so far are listed in table \ref{Aufloesung}. In the present experiment the thermal momentum spread is only about \SI{0.01}{\atomicunit}\ (see chapter \ref{Target properties}) and is, therefore, negligibly small. 

For the electrons the longitudinal momentum resolution is limited by the finite time-of-flight resolution (typically \SI{1}{\nano\second}). For the recoil ions the limiting factors are electric fringe fields and fluctuations of the voltage supplies. In the transverse direction, the finite size of the reaction volume mainly contributes to the uncertainties in the momentum measurements. Applying spatial focusing as discussed in chapter \ref{fieldconfigurations}, can help to significantly improve the target ion transverse resolution. We expect that the recoil ion transverse momentum resolution could even be enhanced by reducing this field, which might be a viable option in experiments where no electrons are detected. 

It should be noted that the momentum resolution does not only depend on the settings of the MOTReMi but also on the projectile beam specifications such as beam diameters, orientations, and pulse durations. In the present experiment, the reaction region is defined by the overlap of the target cloud (\SI{2}{mm} diameter) with the ionizing laser beam which is perpendicular to the spectrometer axis and has a width of about \SI{2}{mm} diameter. In ion impact ionization studies the projectile beam has a diameter of about \SI{1}{mm} and is oriented almost parallel to the spectrometer axis. This allows for somewhat better transverse momentum resolutions for ion impact ionization than for the present photoionization experiment. In general, we expect a further improved resolution in experiments like multi-photon ionization, where the reactions take place in a very small volume with the size of the laser focus (typically a few \SI{10}{\micro\meter}).

\begin{table}
\centering
\begin{tabular}{|l|c|c|}
\hline momentum resolution & transversal & longitudinal \\ 
\hline
\hline Li$^+$ ions& \SI{0.07}{\atomicunit} & \SI{0.035}{\atomicunit} \\ 
electrons & \SI{0.1}{\atomicunit} & \SI{0.01}{\atomicunit} \\ 
\hline \hline
\end{tabular} 
\caption{Resolution (FWHM) of electron and recoil ion momentum components achieved with the MOTReMi.}
\label{Aufloesung}
\end{table}

\subsubsection{Angular emission characteristics}

\begin{figure}
\centering
\includegraphics[angle=0, trim= 2cm 3cm 2cm 3cm, width=0.9\linewidth]{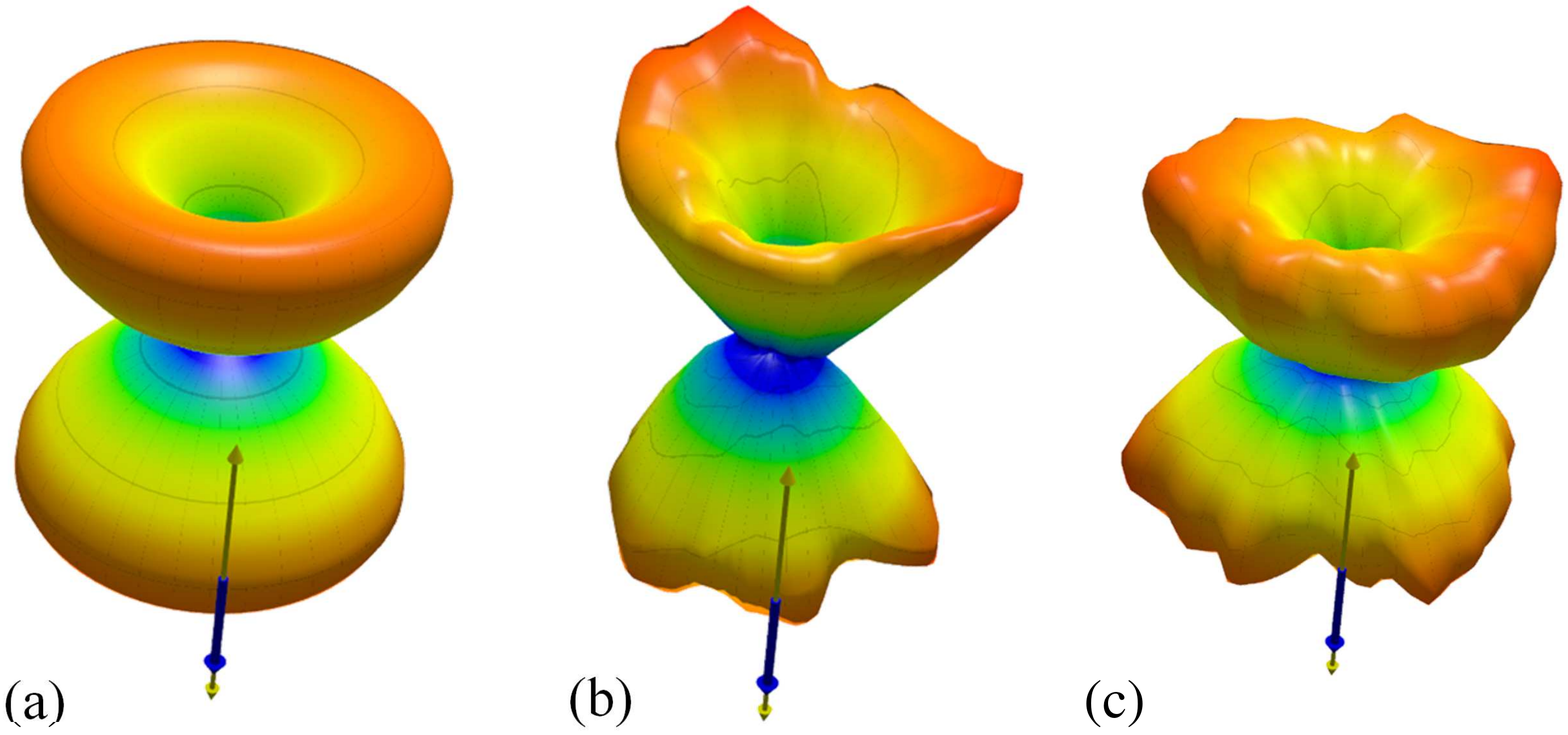}
\caption{Three-dimensional polar plots of angular distributions of the Li fragments after photoionization from the $2^{2}P_{3/2}$ state with $m_L=-1$: 
Absolute square of spherical harmonics $|Y_{2}^{-1}|^{2}$ (a), measured angular distributions of electrons (b) and Li$^+$ ions (c). The yellow and blue arrows indicate the polarizations of photons and target atoms, respectively.}
\label{UVPlot}
\end{figure}

\begin{figure}
\centering
\includegraphics[trim= 3cm 2cm 2cm 0cm, width=0.4\linewidth]{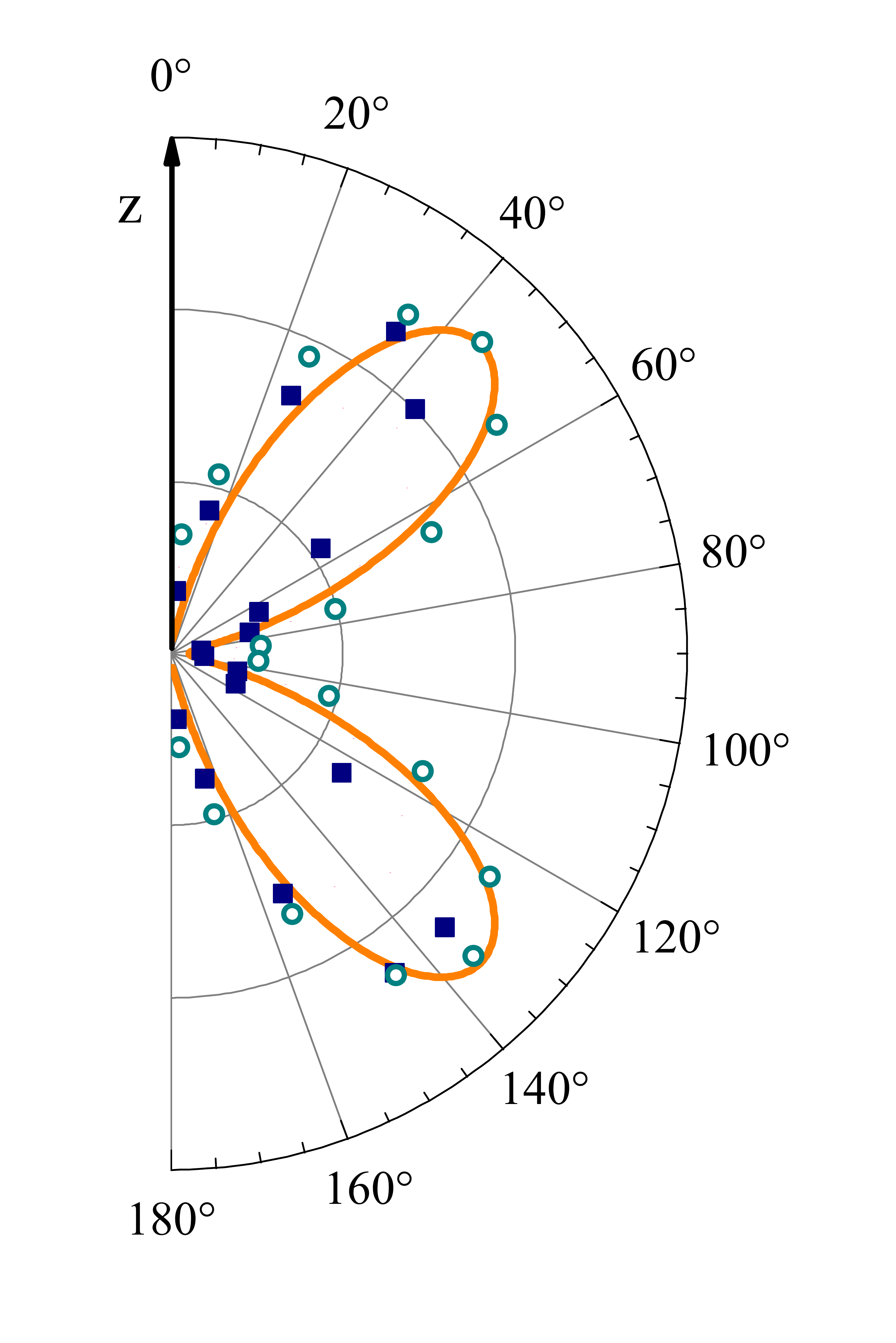}
\caption{Data from Fig.\ \ref{UVPlot} integrated over the azimuthal angle $\varphi$: Spherical harmonics $|Y_{2}^{-1}|^{2}$ (line), electron (full squares) and recoil ion (open circles) $\vartheta$-distributions.}
\label{Kugelflaechenfkt}
\end{figure}

For well-defined laser polarizations and initial target states, general expressions of the angular distribution of emitted electrons are given e.g.\ in Ref.\ \onlinecite{0022-3700-15-6-019}. While the final state is generally a coherent superposition of several partial waves, for some polarization geometries the situation simplifies significantly. Here we discuss the case where the target is initially in the $\left|L=1, m_L=-1\right>$ state and the atom and laser polarizations are aligned to each other. In this case, the ionizing transition fullfills $\Delta m_L=0$ for linear laser polarization, i.e.\ only one partial wave contributes to the final continuum state, namely $\left|L=2, m_L=-1\right>$.

In Fig.\ \ref{UVPlot}(a) the theoretical angular distribution is shown which corresponds to the absolute square of the spherical harmonics $Y_{2,1}(\vartheta,\varphi)$. In (b) and (c) the measured distributions of electron and recoil momenta are plotted, respectively. In general, experimental data and theoretical distribution are in very good qualitative agreement. For the quantitative comparison we integrated the data over $\varphi$ and compared the $\vartheta$ dependence of the cross sections in Fig.\ \ref{Kugelflaechenfkt}. For the measured electron angles the data agrees almost perfectly except in the $\vartheta$ ranges between 45$\degree$ and 65$\degree$ as well as between 115$\degree$ and 135$\degree$. This can be explained by the suppression of the measured electron intensity due to the cyclotron motion of the electrons (cf.\ Fig.\ \ref{pxz} and chapter \ref{wiggles}). The ion data is not compromised by this effect and good agreement is observed.

\subsection{Ion impact ionization}

In the last decade numerous experiments obtaining fully differential cross sections for ion-impact ionization were performed (for a review see Ref.\ \onlinecite{doi:10.1142/S0217751X06032447}). For many of these experiments (and generally for all experiments with reaction microscopes) the electron momentum resolution was strongly impaired in certain parts of the final momentum space due to their cyclotron motion in the homogeneous magnetic field \cite{Buch_Mos}. This limitation has a substantial effect particularly in those experiments where the electric field used for electron extraction is oriented parallel to the projectile beam axis. In the present setup the geometrical configuration and symmetry considerations of the cross sections allow correcting for the above-mentioned constraint and high resolution cross sections can be extracted for almost the full electron momentum space.

\subsubsection{Intensity reconstruction of the final electron momentum space}
\label{wiggles}

\begin{figure}
\centering
\includegraphics[trim= 1cm 2cm 13cm 2cm, width=0.9\linewidth]{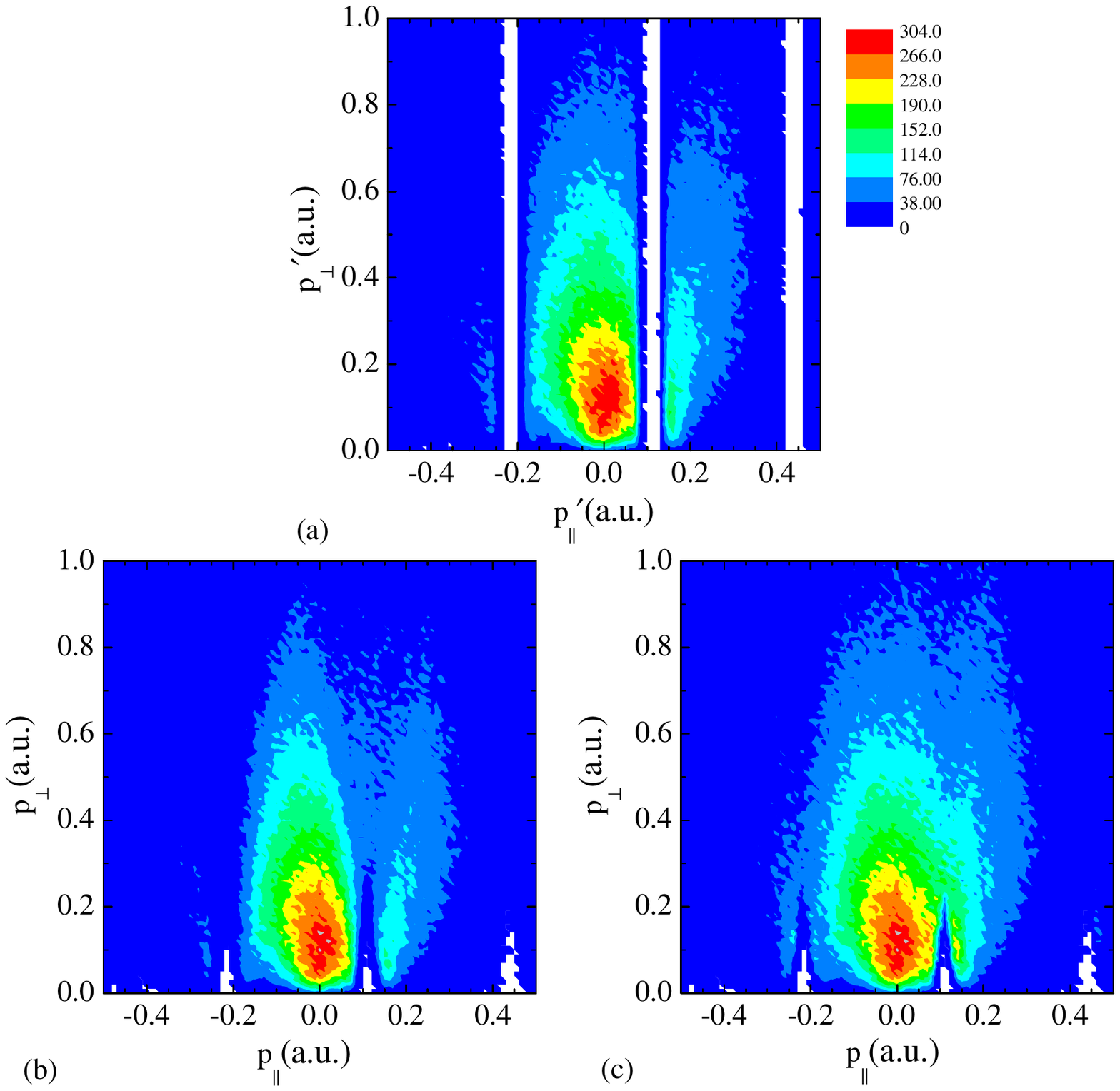}
\caption{Double differential cross sections for the single ionization of lithium in collisions with \SI{16}{\mega\electronvolt} Li$^{2+}$ as a function of the electron momentum components. In (a) the components $p^\prime_\parallel$ and $p^\prime_\perp$ are oriented parallel and perpendicular to the spectrometer axis, respectively. In (b) and (c) the reference frame for $p_\parallel$ and $p_\perp$ is chosen with  respect to the projectile beam axis. In (c) the cross section is shown after the intensity correction described in the text.}
\label{wigglecorrection}
\end{figure}

In reaction microscopes the ejected electrons are guided with parallel electric and magnetic fields onto the position- and time-sensitive detector. While in the longitudinal direction, i.e.\ along the field vectors, the electron motion is only governed by the electric field, the transverse motion is strongly modified by the magnetic field. In the plane perpendicular to the extraction direction the electrons travel on circles whose radii are proportional to their transverse momentum. For certain time-of-flights, when the electrons perform one or several complete turns, they arrive at the same spot on the detector irrespective of their transverse momentum and no energy information is obtained. Those events, i.e.\ with electron time-of-flights close to a multiple of the cyclotron period, are omitted in the analyses resulting in well-defined void ranges in the measured electron momentum spectra. This effect manifests itself in the blank vertical stripes in spectrum Fig.\ \ref{wigglecorrection}(a), where the single ionization cross section in \SI{16}{MeV} $Li^{2+}$ collisions is presented as a function of the electron momenta longitudinal and transverse to the spectrometer axis.

It should be noted that the collision geometry is not tagged by the spectrometer axis. The only relevant orientation defined by the experimental arrangement is the projectile beam direction (at least for unpolarized targets or for target polarizations parallel to the beam axis).  That results in a cylindrical symmetry of the cross section with respect to the direction of the incoming projectile. In other words, the absolute value of the azimuthal angles $\phi$ of the momenta do not contain any important physical information. In contrast, the relative azimuthal angles of e.g.\ electron and recoil ion momenta ($\phi_e-\phi_r$), or electron and outgoing projectile momenta ($\phi_e-\phi_p$) are obviously relevant for the momentum balance in the collisions. Therefore, integrating over $\phi_e$ but preserving the information on the relative angles does not reduce any information on the collision dynamics.

In Fig.\ \ref{wigglecorrection}(b) the same data as in (a) are shown, but the longitudinal direction does not coincide with the spectrometer axis but with the projectile beam. The spectrum is obtained by the $8\degree$ rotation (corresponding to the inclination of the spectrometer axis, see chapter \ref{mechanicaldesign}) of the 3D electron momentum distribution in the spectrometer frame and the integration over $\phi_e$. In this spectrum, the experimentally inaccessible regions are reduced to small triangles close to the abscissa. Moreover, the void regions in the 3D momentum space result in v-shaped areas with suppressed intensity.

The reduction of intensity can be understood as follows: The integration over $\phi_{e}$ corresponds to the summation of the data along a circle in the electron momentum space with radius $p_{\perp}$ centered to the longitudinal axis at the position $p_{\parallel}$.
 Due to the rotation of the reference frame with respect to the spectrometer axis, this circle can partially be located in the void momentum space regions resulting in the reduction of intensity. By which factor the intensity is reduced can easily be derived for each pair of longitudinal and transverse electron momenta. It corresponds to the relative range of $\phi_{e}$ angles for which the abovementioned circle is in the void part of the momentum space. By multiplying the measured intensity with the inverse of this reduction factor the corrected cross section is obtained.

In Fig.\ \ref{wigglecorrection}(c) the cross section is shown after reconstruction of the full intensities. Here only small completely blank regions remain and the v-shaped structures with reduced data almost vanish. Therefore, the present experiment is substantially less affected by resolution issues due to the cyclotron motion of the electrons as compared to earlier experiments where the spectrometer axis was aligned to the projectile beam.

\section{Conclusion}

A reaction microscope (ReMi) combined with a magneto-optically trapped target (MOT) was developed and commissioned. This innovative combination is world-wide unique and it represents in several respects a significant advancement in the investigation of atomic fragmentation dynamics: On the one hand, with alkali and alkaline earth metals a large and very important class of atomic species becomes available as target for scattering experiments. These are particularly appealing, because they have only one (or two, respectively) valence electrons representing rather simple atomic system that allow studying the most fundamental aspects of atomic dynamics. On the other hand, techniques for state-preparation and manipulation of atoms by optical lasers are applicable and the initial target state can be chosen to be excited and even polarized. This provides unique possibilities to study e.g.\ the dependence of the fragmentation dynamics on the initial target state or target orientation.

Experiments on photoionization of laser-excited lithium atoms in \SI{266}{\nano\metre} laser pulses were performed which allow for the detailed characterization of the present setup. The achieved momentum resolution is excellent and for the recoil ions just as well or slightly better than in the best COLTRIMS experiments performed to date. It is presently limited by the spatial extension of the reaction volume and it can be expected to be even improved in experiments with smaller interaction regions. For the preparation of lithium atoms in the excited $2^2P_{3/2}$ state a polarization of close to \SI{100}{\percent} was achieved by employing optical pumping.

The MOTReMi represents a first step in applying the comprehensive toolbox developed for the preparation and manipulation of atomic quantum-gases to scattering experiments. In future experiments, optical trapping, cooling (even to degeneracy) and manipulation will allow for an unprecedented level of target control providing insights in the correlated dynamics of few-particle quantum-systems on a hitherto unreachable level of detail.

\begin{acknowledgments}

We are indebted to S.~Jochim, A.~Dorn, and R.~ Moshammer and their groups for many helpful discussions. This work was funded by the German Research Council (DFG), under Grant No. FI 1593/1-1. We are grateful for the support from the Alliance Program of the Helmholtz Association (HA216/EMMI). V.L.B.J thanks for the support from the DAAD and CAPES. M.S. acknowledges support from the National Science Foundation, under Grant No. 1401586.

\end{acknowledgments}

\def\urlprefix{}
\def\url#1{}
\bibliographystyle{apsrev}
\bibliography{Literaturverzeichnis}

\end{document}